\documentclass[sigconf,10pt]{acmart}
\usepackage{graphicx}
\usepackage{balance}  % for  \balance command ON LAST PAGE  (only there!)

\usepackage{booktabs} % For formal tables
\usepackage{etoolbox}
\usepackage{xspace,colortbl}
\usepackage{datetime}
\usepackage{tabularx}
\usepackage{fancyvrb}
\usepackage{enumitem}
\usepackage[ruled,vlined,linesnumbered]{algorithm2e}
\usepackage{letltxmacro}
\usepackage{url}
\usepackage{mathtools}
\usepackage{placeins}
\usepackage{dpfloat}
\usepackage{listings}
\newif\ifarxiv
\arxivtrue  % comment out for arxiv submission

\definecolor{Gray}{gray}{0.9}
\newcommand{\cf}{\cellcolor{Gray}\bf}

\usepackage{breakurl}

\newcommand{\smallsec}[1]{\vspace{0.3em}\noindent{\bf #1}}
\newcommand{\sys}{{\sc ExplainIt!}\xspace}
\newcommand{\squeeze}{\vspace{0em}}
\newcommand{\X}{\mathbf{X}}
\newcommand{\Y}{\mathbf{Y}}
\newcommand{\Z}{\mathbf{Z}}
\newcommand{\BigO}{\mathcal{O}}
\LetLtxMacro\latexincludegraphics\includegraphics
\renewcommand{\includegraphics}[2][]{%
  \latexincludegraphics[#1]{#2}\squeeze}
\DeclarePairedDelimiter\abs{\lvert}{\rvert}%

\usepackage{dcolumn}
\newcolumntype{d}[1]{D{.}{.}{#1}}

\copyrightyear{2019}
\acmYear{2019}
\setcopyright{acmcopyright}
\acmConference[SIGMOD '19]{2019 International Conference on Management of Data}{June 30-July 5, 2019}{Amsterdam, Netherlands}
\acmBooktitle{2019 International Conference on Management of Data (SIGMOD '19), June 30-July 5, 2019, Amsterdam, Netherlands}
\acmPrice{15.00}
\acmDOI{10.1145/3299869.3314048}
\acmISBN{978-1-4503-5643-5/19/06}

\renewcommand\footnotetextcopyrightpermission[1]{} % removes footnote with conference information in first column
\begin{document}

% ****************** TITLE ****************************************

\title{\sys -- A declarative root-cause analysis engine for time
  series data\ifarxiv~(extended version)\fi}

\author{Vimalkumar Jeyakumar}
\affiliation{Cisco Tetration Analytics}
\email{jvimal@tetrationanalytics.com}

\author{Omid Madani}
\affiliation{Cisco Tetration Analytics}
\email{omadani@tetrationanalytics.com}

\author{Ali Parandeh}
\affiliation{Cisco Tetration Analytics}
\email{aparande@tetrationanalytics.com}

\author{Ashutosh Kulshreshtha}
\affiliation{Cisco Tetration Analytics}
\email{ashutkul@tetrationanalytics.com}

\author{Weifei Zeng}
\affiliation{Cisco Tetration Analytics}
\email{weifzeng@tetrationanalytics.com}

\author{Navindra Yadav}
\affiliation{Cisco Tetration Analytics}
\email{nyadav@tetrationanalytics.com}
%\end{comment}
% There's nothing stopping you putting the seventh, eighth, etc.
% author on the opening page (as the 'third row') but we ask,
% for aesthetic reasons that you place these 'additional authors'
% in the \additional authors block, viz.
\date{\today}
% Just remember to make sure that the TOTAL number of authors
% is the number that will appear on the first page PLUS the
% number that will appear in the \additionalauthors section.
\renewcommand{\shortauthors}{V. Jeyakumar et al.}

\begin{abstract}
  We present \sys, a declarative, unsupervised root-cause analysis
  engine that uses time series monitoring data from large complex
  systems such as data centres.  \sys empowers operators to succinctly
  specify a large number of causal hypotheses to search for causes of
  interesting events.  \sys then ranks these hypotheses, reducing the
  number of causal dependencies from hundreds of thousands to a
  handful for human understanding.  We show how a declarative
  language, such as SQL, can be effective in declaratively enumerating
  hypotheses that probe the structure of an unknown probabilistic
  graphical causal model of the underlying system.  Our thesis is that
  databases are in a unique position to enable users to rapidly
  explore the possible causal mechanisms in data collected from
  diverse sources.  We empirically demonstrate how \sys had helped us
  resolve over 30~performance issues in a commercial product since
  late 2014, of which we discuss a few cases in detail.
\end{abstract}

\maketitle

\section{Introduction}\label{sec:intro}
% Explain the general environment
In domains such as data centres, econometrics~\cite{fred}, finance,
systems biology~\cite{seth2015granger}, and many others~\cite{nasa},
there is an explosion of time series data from monitoring complex
systems.  For instance, our product {\em Tetration Analytics} is a
server and network monitoring appliance, which collects millions of
observations every second across tens of thousands of servers at our
customers.  Tetration Analytics itself consists of hundreds
of services that are monitored every minute.
% Need other key examples from the community in this machine
% monitoring space.

% What's the objective?  Example links to external cases.
% Explain some behaviour that happened in the past.
% in our experience: not necessarily an anomaly.
% Always interested in events that are happening or in the past.
% At various levels of abstraction.
One reason for continuous monitoring is to understand the dynamics of
the underlying system for root-cause analysis.  For instance, if a
server's response latency shows a spike and triggered an alert,
knowing what caused the behaviour can help prevent such alerts from
triggering in the future.  In our experience debugging our own
product, we find that root-cause analysis (RCA) happens at various
levels of abstraction mirroring team responsibilities and
dependencies: an operator is concerned about an affected service, the
infrastructure team is concerned about the disk and network
performance, and a development team is concerned about their
application code.

% Strawman approaches and why those are insufficient.
% What's unique to this setting?
To help RCA, many tools allow users to query and classify
anomalies~\cite{bailis2017macrobase}, correlations between pairs of
variables~\cite{pelkonen2015gorilla,vmware-wavefront}.  We find that
the approaches taken by these tools can be unified in a single
framework---causal probabilistic graphical
models~\cite{pearl2009causality}.  This unification permits us to
generalise these tools to more complex scenarios, apply various
optimisations, and address some common issues:
\begin{itemize}[noitemsep,leftmargin=1em,topsep=1pt]
\item {\bf Dealing with spurious correlations}: It is not uncommon to
  have per-minute data, yet hundreds of thousands of time series.  In
  this regime the number of data points over even \emph{days} is in
  the thousands, and is at least two orders of magnitude fewer than
  the dimensionality (hundreds of thousands).  It is no surprise that
  one can always find a correlation if one looks at enough data.
\item {\bf Addressing specificity}: Some metrics have trend and
  seasonality (i.e., patterns correlated with time).  It is important
  to have a principled way to remove such variations and focus on
  events that are interesting to the user, such as a spike in
  latency~\S\ref{subsec:pseudocauses}.
\item {\bf Generating concise summaries}: We firmly believe that
  summarising into human-relatable groups is key to scale
  understanding~\S\ref{subsec:featurefamilies}.  Thus, it becomes
  important to organise time series into {\em groups}---dynamically
  determined at users' direction---and rank the candidate causes
  between groups of variables in a theoretically sound way.
\end{itemize}

We created \sys, a large-scale root-cause inference engine and
explicitly addressed the above issues.  \sys is based on three
principles: First, \sys is designed to put humans in the loop by
exposing a {\em declarative} interface (using SQL) to {\em
  interactively} query for explanations of an observed phenomena.
Second, \sys exploits side-information available in time series
databases (metric names and key-value annotations) to enable the user
to group metrics into meaningful {\em families} of variables.  And
finally, \sys takes a {\em principled approach} to rank candidate
families (i.e., ``explanations'') using causal data mining techniques
from observational data.  \sys ranks these families by their {\em
  causal relevance} to the observed phenomenon in a {\em
  model-agnostic} way.  We use statistical dependence as a yardstick
to measure causal relevance, taking care to address spurious
correlations.

% Sneak peak...
We have been developing \sys to help us diagnose and fix performance
issues in our product.  A key distinguishing aspect of \sys is that it
takes an {\em ab-initio} approach to help users uncover interactions
between system components by making as few assumptions as necessary,
which helps us be broadly applicable to diverse scenarios.  The user
workflow consists of three steps: In step~1, the user selects both the
target metric and a time range they are interested in.  In step~2, the
user selects the search space among all possible causes.  Finally in
step~3, \sys presents the user with a set of candidate causes ranked
by their predictability.  Steps 2--3 are repeated as needed.  (See
Figure~\ref{fig:explainit-webapp} in Appendix for prototype
screenshots.)

\smallsec{Key contributions}: We substantially expand on our earlier
work~\cite{explainit-causalml2018} and show how database systems are
in a unique position to accomplish the goal of exploratory causal data
analysis by enabling users to declaratively enumerate and test causal
hypotheses.  To this end:
\begin{itemize}[noitemsep,leftmargin=1em,topsep=1pt]
  \item We outline a design and implementation of a pipeline using a
    unified causal analysis framework for time series data at a large
    scale using principled techniques from probabilistic graphical
    models for causal inference (\S\ref{sec:approach}).
  \item We propose a ranking-based approach to summarise dependencies
    across variables identified by the user
    (\S\ref{sec:impl}).
  \item We share our experience troubleshooting many real world
    incidents (\S\ref{sec:case-study}): In over 44~incidents spanning
    4~years, we find that \sys helped us satisfactorily identify
    metrics that pointed to the root-cause for 31~incidents in {\em
      tens of minutes}.  In the remaining 13~incidents, we could not
    diagnose the issue because of insufficient monitoring.
  \item We evaluate concrete ranking algorithms and show why a single
    ranking algorithm need not always work (\S\ref{sec:evaluation}).
\end{itemize}
\ifarxiv \else We discuss one more case study, lessons learnt, and
additional proofs in our extended version of our
paper~\cite{explainit-extended}.\fi

Although correlation does not imply causation, having humans in the
loop of causal discovery~\cite{tenenbaum2003theory} side-steps many
theoretical challenges in causal discovery from observational
data~\cite[Chap.~3]{pearl2009causality}.  Furthermore, we find that a
declarative approach enables users to both generate plausible
explanations among all possible metric families, or confirm hypotheses
by posing a targeted query.  We posit that the techniques in \sys are
generalisable to other systems where there is an abundance of time
series organised hierarchically.
 % 2pg
\section{Background}
% Examples of complex systems environment
We begin by describing a familiar target environment for \sys, where
there is an abundance of machine-generated time series data: data
centres.  Various aspects of data centres, from infrastructure such as
compute, memory, disk, network, to applications and services' key
performance metrics, are continuously monitored for operational
reasons.  The scale of ingested data is staggering: Twitter/LinkedIn
report over 1~billion metrics/minute of data.  On our own deployments,
we see over 100~Million flow observations every minute across tens of
thousands of machines, with each observation collecting tens of
features per flow.

In these environments time series data is structured: An
event/observation has an associated timestamp, a list of key-value
categorical attributes, and a key-value list of numerical
measurements.  For example: The network activity between two hosts
{\small\tt datanode-1} and {\small\tt datanode-2} can be represented
as:
\begin{Verbatim}[fontsize=\small]
  timestamp=0
  flow{src=datanode-1, dest=datanode-2,
       srcport=100, destport=200, protocol=TCP}
  bytecount=1000 packetcount=10 retransmits=1
\end{Verbatim}
Here, the tag keys are {\small\tt src}, {\small\tt dest} and
{\small\tt srcport, destport} joined with three measurements
({\small\tt bytecount}, {\small\tt packetcount}, and {\small\tt
  retransmits}).  Such representations are commonly used in many time
series database and analytics tools~\cite{opentsdb,yang2014druid}.
Throughout this paper, when we use the term {\em metric} we refer to a
one-dimensional time series; the above example is three-dimensional.

\begin{figure}[t]
\centering
\includegraphics[width=0.8\columnwidth]{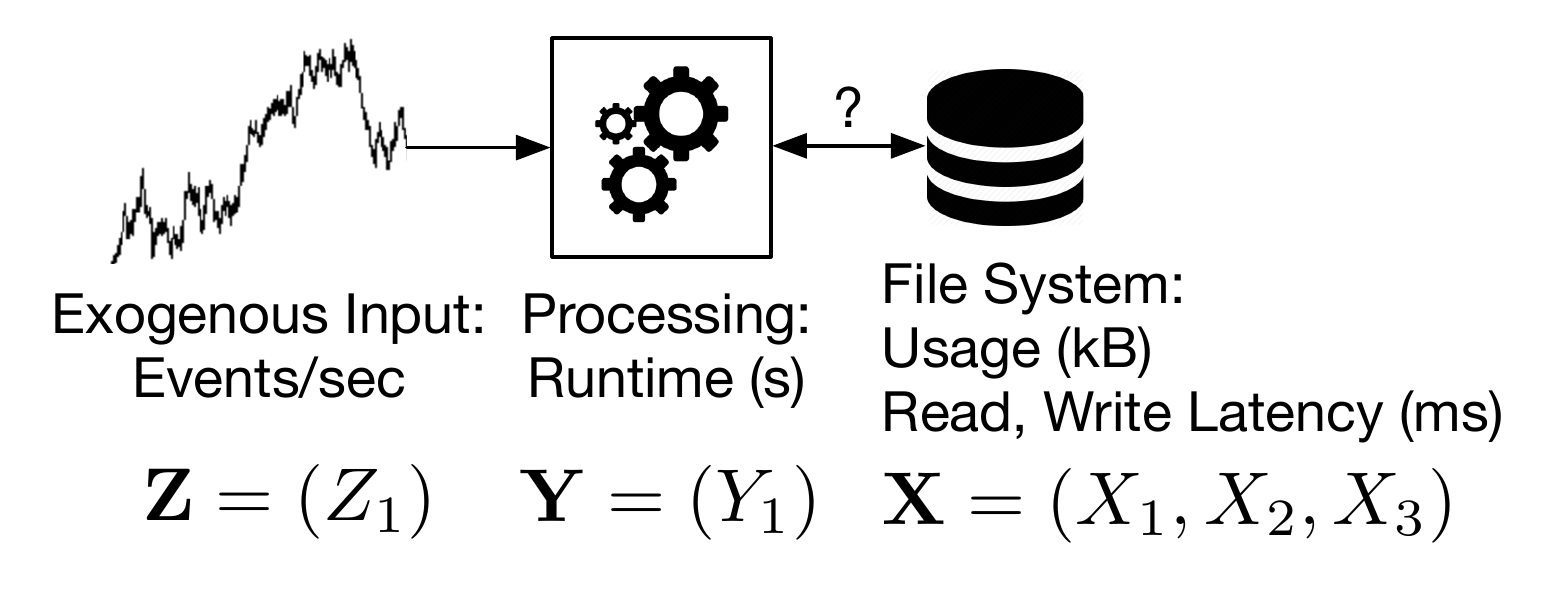}
\caption{A simplified representation of a data processing pipeline,
  whose five performance indicators $(X_1,\ldots,Y_1,Z_1)$ can be used
  by \sys for offline analysis.  It is plausible that a high runtime,
  due to a large data output, could result in a higher disk latency.
  The reverse causal relationship is also plausible: a rogue service
  trashing disk performance could affect the pipeline's runtime.}
\label{fig:simple-example}\squeeze
\end{figure}

\section{Approach}\label{sec:approach}
To illustrate our approach we will use an application shown in
Figure~\ref{fig:simple-example}: a real-time data processing pipeline
with three components that are monitored.  First, the input to the
system is an event stream whose input rate events/second is the time
series $\Z(t) = (Z_1(t))$.  The second component is a pipeline that
produces summaries of input, and its average processing time per
minute is $\Y(t)=(Y_1(t))$.  Finally, the pipeline outputs its result
to a file system, whose disk usage $X_1$ and average read/write
latency $X_2,X_3$ are collectively grouped into $\X(t)=(X_1(t),
X_2(t), X_3(t))$.  For brevity, we will drop the time $t$ from the
above notations.  Thus, in this example our system state is captured
by the set of variables $(\X, \Y, \Z)$.

\begin{figure}[t]
\centering
\includegraphics[width=0.32\textwidth]{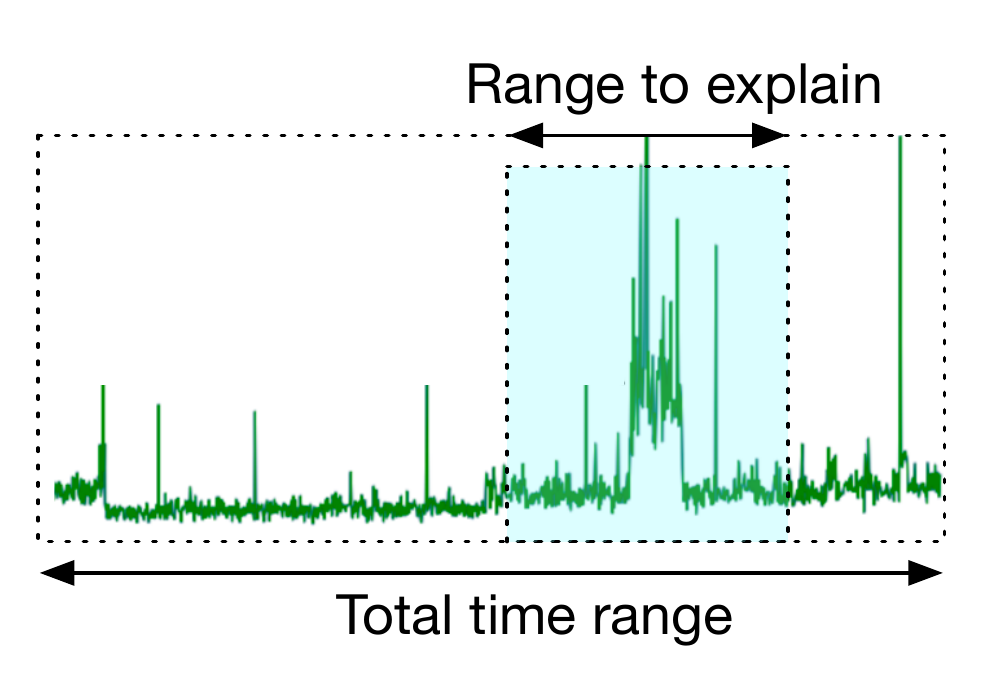}
\caption{Each scenario requires the user to specify two time ranges: A
  total time range (e.g., last 1 day), and a time range of a specific
  event that the user wishes to be explained.}
\label{fig:time-range}
\end{figure}

% Put user in the loop: Specify hypotheses
\smallsec{Workflow}: As mentioned in \S\ref{sec:intro}, we require the
users to specify the target metric(s) of interest (denoted by $\Y$).
Typically, these are key performance indicators of the system.  Then,
users specify two time ranges: one that roughly includes the overall
time horizon (typically, a few days of minutely data points are
sufficient for learning), and another (optional) overlapping time
range to highlight the performance issue that they are interested in
root-causing (see Figure~\ref{fig:time-range}).  If the second time
range is not specified, we default to the overall time range.  In this
step, the user also specifies a list of metrics to control for
specificity (denoted by $\Z$), as described in
\S\ref{subsec:pseudocauses}.  Finally, the user specifies a search
space of metrics (denoted by $\X$) that they wish to explore using
SQL's relational operators.  \sys scores each hypothesis in the search
space and presents them in the order of decreasing scores (with a
default limit of top 20) to the user (\S\ref{subsec:impl:scoring}).
The user can then inspect each result, and fork off further analyses
and drill down to narrow the root-cause.
Algorithm~\ref{alg:explainit} is the pseudocode to \sys's main
interactive search loop.

Due to its ab-initio approach, \sys is only typically used when the
usual processes in place such as monitoring dashbords, rules, or
alerts are insufficient.  After a typical session in \sys, the user
identifies a small set of metrics that are useful for frequent
diagnosis to create new dashboards and alerts.

% Figure for
%% Input data: Any source that can be transformed to this schema:
%% Timestamp, metric name, dimensions (KV pairs), metrics (KV pairs)

%% Featurisation:
%% Transform input to Timestamp, group, instance, metric

%% Scoring input:
%% - List of feature families.
%% - Target feature family.
%% - Conditioning features.
%% - Options: key-value pairs

%% Output explanation:
%% name, group name, score, debug key-value pairs.
\subsection{Model for hypotheses}
For a principled approach to root-cause analysis, we found it helpful
to view each underlying metric as a node in some unknown causal
Bayesian Network (BN)~\cite{pearl2009causality}.  A BN is a directed
acyclic graph (DAG) in which the nodes are random variables, and the
graph structure encodes a set of probabilistic conditional
dependencies: Each variable is conditionally independent of its
non-descendants given its parents~\cite{pearl2009causality}.  In a
causal BN the directed edges encode cause-effect relationship between
the variables; that is, the edge $\Z \rightarrow \Y$ encodes the fact
that $\Z$ {\em causes} $\Y$.  Put another way, an intervention in $\Z$
(e.g., higher/lower input events) results in a change in the
distribution of $\Y$ (higher/lower average processing time), but an
intervention in $\Y$ (e.g., artificially slowing down the pipeline)
does not affect the distribution of $\Z$.  One possible causal
hypothesis for the dynamics of the example is (a)~the chain:
$\Z\rightarrow\Y\rightarrow\X$ or $\Z\leftarrow\Y\leftarrow\X$; other
hypotheses are (b)~the fork: $\Y\leftarrow\Z\rightarrow\X$ and (c) the
collider: $\Y\rightarrow\Z\leftarrow\X$.

\begin{algorithm}[t]
  \SetKwData{SearchFamilies}{SearchFamilies}
  \SetKwData{ConditioningFamilies}{$\Z$} \KwData{Metric names,
    key-value attributes, time series} \KwIn{Target metric (or family)
    $\Y$} \While{user not satisfied}{ \SearchFamilies$\leftarrow$
    \emph{All families or user defined subset}\;
    \ConditioningFamilies$\leftarrow$ \emph{$\varnothing$ or user
      defined subset to condition or pseudocause derived from $\Y$}\;
    \lForEach{family $\X_i \in {\rm \SearchFamilies \;except\;} \Y,
      \Z$}{
       \label{alg:explainit:parallelstep}
       \emph{``assoc'' returns a value between 0 (low score) and 1
         (high score) for the dependence $\Y \sim \X_i \mid \Z$}
       ${\rm score}(\X_i) \leftarrow {\rm assoc}(\Y, \X_i \mid \Z)$\;}
     \emph{Show $\X_i$'s to user sorted by decreasing {\rm score}($\X_i$)}\;
  }
  \caption{Pseudocode for the core ranking and interactive loop in
    \sys.  Naturally, once the users review the results they can pose
    additional queries to further narrow down the candidate metrics of
    interest.}
\label{alg:explainit}
\end{algorithm}

The root-cause analysis problem translates to finding only the {\em
  ancestors} of a key set of variables ($\Y$) that measure the
observed phenomenon, in DAG structures that encode the same
conditional dependencies as seen in observations from the underlying
system.  In our experience, we neither needed to learn the full
structure between all variables, nor the actual parameters of the
conditional dependencies in the BN.

The causal BN model makes the following assumptions:
\begin{itemize}[noitemsep,leftmargin=1em,topsep=0pt]
\item Causal Markov / Principle of Common Cause: Any observed
  dependency (measured by say the correlation) between variables
  reflect some structure in the DAG~\cite{sep-physics-Rpcc}.  That is,
  if $\X$ is not independent of $\Y$ (i.e. $\X \not\perp \Y$), then
  $\X$ and $\Y$ are connected in the graph.
\item Causal Faithfulness: The structure of the graph implies
  conditional independencies in the data.  For the example in
  Figure~\ref{fig:simple-example} the causal hypothesis $\Z\rightarrow
  \Y\rightarrow \X$ implies that $\Z \perp \X \mid \Y$.
\end{itemize}

Taken together, these assumptions help us infer that (a)~the existence
of a dependency between observed variables $\X$ and $\Y$ mean that
they are connected in the graph formed by replacing the directed edges
with undirected edges; and (b)~the {\em absence} of dependency between
$\X$ and $\Y$ in the data mean there is no causal link between them.
The assumptions are discussed further in the book
Causality~\cite[Sec.~2.9]{pearl2009causality}.

%% \footnote{Another
%%   assumption, called the {\em Causal sufficiency} assumption states
%%   that the set of measured metrics include all common causes of pairs
%%   of variables.  The trend towards monitoring every aspect of computer
%%   systems means that we are unlikely to miss monitoring critical
%%   variables for troubleshooting.}

\smallsec{Why?}  The above approach offers three main benefits.
First, the formalism is a non-parametric and {\em declarative} way of
expressing dependencies between variables and defers any specific
approach to the runtime system.  Second, the unified approach
naturally lends itself to multivariate dependencies of more complex
relationships beyond simple correlations between pairwise univariate
metrics.  Third, the approach also gives us a way to reason about
dependencies that might be easier to detect only when holding some
variables constant; see conditioning/pseudocauses
(\S\ref{subsec:pseudocauses}) for an example and explanation.

Each of these reasons informs \sys's design: The declarative approach
can be used to succinctly express a large number of candidate
hypotheses for both univariate and multivariate cases.  We also show
how {\em conditional probabilities} can be used to search explanations
for specific variations in the target variable, improving overall
ranking.

\subsection{Feature Families}\label{subsec:featurefamilies}
Grouping univariate metrics into families is useful to reduce the
complexity of interpreting dependencies between variables.  Hence,
grouping is a critical operation that precedes hypothesis generation.
Each metric has tags that can be used to group; for example, consider
the following metrics:
\begin{table}[h]
  \centering\small
  \begin{tabular}{l|l}
    {\bf Name} & {\bf Tags} \\\hline
    {\tt input\_rate} & {\tt type=event-1} \\\hline
    {\tt input\_rate} & {\tt type=event-2} \\\hline
    {\tt runtime} & {\tt component=pipeline-1} \\\hline
    {\tt disk}  & {\tt host=datanode-1, type=read\_latency} \\\hline
    {\tt disk} & {\tt host=datanode-2, type=read\_latency} \\\hline
    {\tt disk} & {\tt host=namenode-1, type=read\_latency} \\\hline
  \end{tabular}
\end{table}

We can group metrics their name, which gives us three hypotheses:
{\small\tt input\_rate\{*\}, runtime\{*\}, disk\{*\}}.  Or, we can
group the metrics by their host attribute, which gives us four
families:
\begin{Verbatim}[fontsize=\small]
  *{host=datanode-1}, *{host=datanode-2},
  *{host=namenode-1}, *{host=NULL}
\end{Verbatim}

The first family captures all metrics on host datanode-1, can be used
to create a hypothesis of the form ``Does {\em any} activity in
datanode-1 \ldots?''  Using SQL, users also have the flexibility to
group by a pattern such as {\small\tt disk\{host=datanode*\}}, which
can be used to create a hypothesis of the form ``Does {\em any}
activity in {\em any} datanode \ldots?''  They can incorporate other
meta-data to apply even more restrictions.  For example, if the users
have a machine database that tracks the OS version for each hostname,
users can join on the hostname key and select hosts that have a
specific OS version installed.  We list many example queries in
Appendix~\ref{sec:sql-queries}.

\subsection{Generating hypotheses}
A causal hypothesis is a triple of feature families $(\X,\Y,\Z)$,
organised as (a) an explainable feature---$\X$, (b) the target
variable---$\Y$, and (c) another list of metrics to condition
on---$\Z$.  Clearly, there should be no overlap in metrics between
$\X$, $\Y$ and $\Z$.  While $\X$ and $\Y$ must contain at least one
metric, $\Z$ could be empty.  Testing any form of dependency (chains,
forks, or colliders) in the causal BN can be reduced to scoring a
hypothesis for appropriate choices of $\X,\Y,\Z$; see the PC algorithm
for more details~\cite{spirtes2000causation}.  While one could
automatically generating exponentially many hypotheses for all
possible groupings, we rely on the user to constrain the search space
using domain knowledge.

The hypothesis specification is guided by the nature of exploratory
questions focusing on subsystems of the original system.  In
Figure~\ref{fig:simple-example}, this would mean: ``does some activity
in the file system $\X$ explain the increase in pipeline runtimes $\Y$
that is not accounted for by an increase in input size $\Z$?''
Contrast this to a very specific (atypical) query such as, ``does disk
utilisation on server~1 explain the increase in pipeline runtime?''
We can operationalise the questions by converting them into
probabilistic dependencies: The first question asks whether $\X \perp
\Y \mid \Z$.  We can evaluate this by testing whether $\Y$ is
conditionally independent of $\X$ given $\Z$, i.e., whether $P(\Y \mid
\X, \Z) = P(\Y \mid \Z)$ (\S\ref{subsec:impl:scoring}).

\subsection{Conditioning and pseudocauses}\label{subsec:pseudocauses}
The framework of causal BNs also help the user focus on a specific
variation pattern inherent in the data in the presence of multiple
confounding variations.  Consider a scenario in which $Y_1$ (in
Figure~\ref{fig:simple-example}) has two sources of variation: a
seasonal component $Y_s$ and a residual spike $Y_r$, and the user is
interested in explaining $Y_r$.
\begin{figure}[t]
\centering
\includegraphics[width=0.4\textwidth]{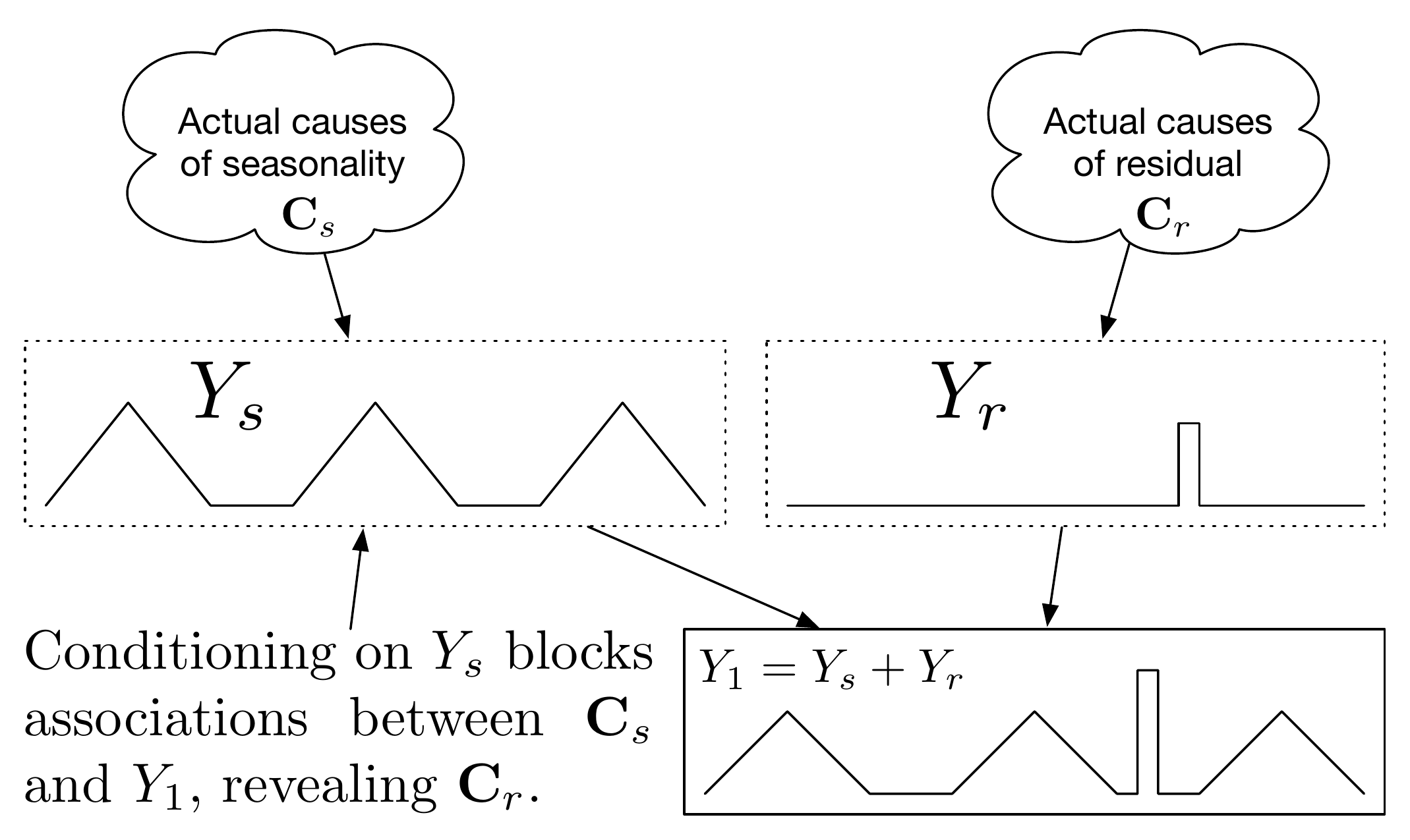}
\caption{Conceptual Bayes Network to illustrate pseudocauses that can
  be derived from decomposing $Y_1$ into its constituent parts.
  Conditioning on $Y_s$ is an optimisation that allows us to boost
  $\mathbf{C}_r$'s ranking without having to find $\mathbf{C}_s$.}
\label{fig:pseudocause}
\end{figure}

We can conceptualise this problem using the causal BN shown in
Figure~\ref{fig:pseudocause} under the assumption that there are
independent causes for $Y_r$ and $Y_s$.  By conditioning on the causes
of $Y_s$, we can find variables that are correlated {\em only with}
$Y_r$ and not with $Y_s$, which helps us find specific causes of
$Y_r$.

However, we often run into scenarios where the user does not know or
is not interested in finding what caused $Y_s$ (i.e., the parents of
$Y_s$).  The causal BN shown in Figure~\ref{fig:pseudocause} offers an
immediate graphical solution: to explain $Y_r$, it is sufficient to
condition on the pseudocause $Y_s$ (derived from $Y$) to ``block'' the
effect of the true causes of seasonality ($\mathbf{C}_s$) without
finding it.  Although prior work~\cite{bailis2017macrobase} has shown
how to express such transformations (trend identification,
seasonality, etc.)  our emphasis here is to show how techniques from
causal inference offer a principled way of {\em reasoning} about such
optimisations, helping \sys generate explanations specific to the
variation the user is interested in.
% data -> select(target) -> trend_seasonality -> {target_trend,
% target_seasonality, target_residual}

% Scoring: D = data union target_processed
% Z  = D -> select(target_seasonality, target_trend)
% Xi = D -> select(*) except target*
% Y  = D -> target_residual

%Appendix~\ref{} discusses more examples.
% example system
%% Data:
% Target
% Inputs
% Conditioning variables
% Schema:

\subsection{Hypothesis ranking}\label{subsec:impl:scoring}
Recall that scoring a hypothesis triple $(\X,\Y,\Z)$ quantifies the
degree of dependence between $\X$ and $\Y$ controlling for $\Z$.  Each
element of the triple contains one or more univariate variables.  We
implemented several scoring functions that can be broadly classified
into (a)~univariate scoring to only look at marginal dependencies
(when $\Z=\varnothing$), and (b)~multivariate scoring to account for
joint dependencies.

\begin{figure}[t]
\centering
\includegraphics[width=0.5\textwidth]{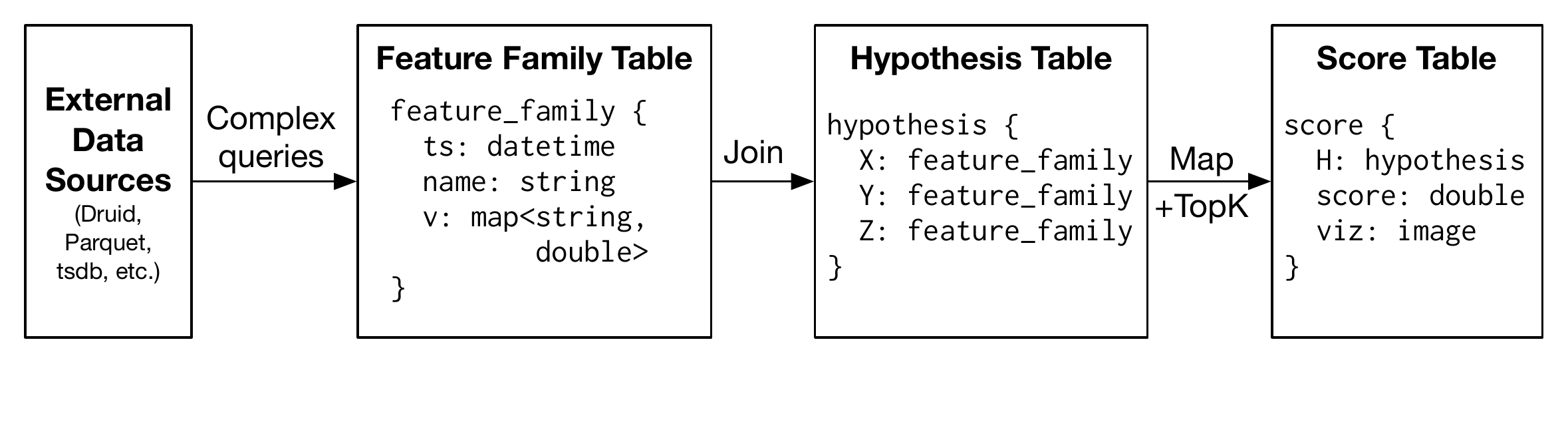}
\caption{\sys's end-to-end pipeline combining complex event parsing
  and extraction in the first stage to generate and score hypotheses.}
\label{fig:pipeline}
\end{figure}

\smallsec{Univariate scoring}: When $\Z=\varnothing$, we can summarise
the dependency between $\X$ and $\Y$ by first computing the matrix of
Pearson product-moment correlation $\rho_{ij}$ between each univariate
element $X_i \in \X$ and $Y_j \in \Y$.
\begin{comment}
\[
\rho_{i j} = \frac
  {\sum\limits_{t=1}^T\left(X_i(t)-\bar{X}_i\right)\left(Y_j(t)-\bar{Y}_j\right)}
  {\sqrt{\sum\limits_{t=1}^T\left(X_i(t)-\bar{X}_i\right)^2}\sqrt{\sum\limits_{t=1}^T\left(Y_j(t)-\bar{Y}_j\right)^2}}
\]
where $\bar{X}_i=\frac{1}{T}\sum_tX_i(t)$, and
$\bar{Y}_j=\frac{1}{T}\sum_tY_j(t)$.
\end{comment}
To summarise the dependency into a single score, we can either compute
the average or the maximum of their absolute values:
\begin{align*}
  {\rm CorrMean} &= \underset{ij}{\rm mean \;} \abs{\rho_{ij}} \\
   {\rm CorrMax} &= \underset{ij}{\max \;} \abs{\rho_{ij}}
\end{align*}

When $\Z$ is non-empty, we use the scoring mechanism outlined below
that unifies joint and conditional scoring into a single method.

\smallsec{Multivariate and conditional scoring}: To handle more
complex hypothesis scoring, we seek to derive a single number that
quantifies to what extent $\X \sim \Y \mid \Z$.  When $\Z =
\varnothing$, we perform a regression where the input data points are
from the same time instant, i.e. $(\X(t), \Y(t))$.  \footnote{The user
  could specify lagged features from the past when preparing the input
  data (by using {\tt LAG} function in SQL).}  One could use
non-linear regression techniques such as spline regression, or neural
networks, but we empirically found that linear regression is
sufficient.  The regression minimises mean squared loss function $L$
between the predicted $\hat{\Y}$ and the observed $\Y$ over $T$ data
points. After training the model, we compute the prediction
$\hat{\Y}$, and the residual $\mathbf{R}_{\Y;\X} = \Y - \hat{\Y}$,
which is the ``unexplained'' component in $\Y$ after regressing on
$\X$.  The variance in this residual {\em relative} to the variance in
the original signal $\Y$ (call it $1-r_{\Y;\X}^2$) varies between 0
($\X$ perfectly predicts $\Y$) and 1 ($\X$ does not predict $\Y$).
The score is this value $r^2$.
%% \[
%% L(\hat{\Y},\Y) = \frac{1}{T}\sum\limits_{t=1}^T \left(\hat{\Y}(t) - \Y(t)\right)^2
%% \]

When $\Z$ is not empty, we require multiple regressions.  First, we
regress $\Y \sim \Z$ to compute the residuals $\mathbf{R}_{\Y;\X}$.
Similarly, we regress $\X \sim \Z$ to compute the residual
$\mathbf{R}_{\X;\Z}$.  Finally, we regress $\mathbf{R}_{\Y;\Z} \sim
\mathbf{R}_{\X;\Z}$ and compute the percentage of variance
$r_{\Y;\X\mid\Z}^2$ in the residual $\mathbf{R}_{\Y;\Z}$ explained by
$\mathbf{R}_{\X;\Z}$ as outlined above.  This percentage of variance
is conditional on $\Z$; intuitively, if the score (percent variance
explained) is high, it means that there is still some residual in $\Y
\mid \Z$ that can be explained by $\X \mid \Z$, which means that $\Y
\not\perp \X \mid \Z$.  If $\X$, $\Y$, and $\Z$ are jointly normally
distributed, and the regressions are all ordinary least squares, then
one can show that the above procedure gives a zero conditional score
iff $\X \perp \Y \mid \Z$.  The proof is in the appendix of the
extended version of this paper~\cite{explainit-extended}.

The score obtained by the above procedure has an overfitting problem
when we have a large number of predictors in $\X$ and a small number
of observations.  To combat this, we use two standard techniques:
First, we apply a penalty (we experimented with both $\mathrm{L}_1$
penalty (Lasso) and $\mathrm{L}_2$ penalty (Ridge)) on the
coefficients of the linear regression.  Second, we use $k$-fold
cross-validation for model selection (with $k=5$), which ensures that
the $r^2$ score is an estimate of the model performance on unseen
data~(also called the {\em adjusted} $r^2$; see
Appendix~\ref{sec:r2}).  Since we are dealing with time series data
that has rich auto-correlation, we ensure that the validation set's
time range does not overlap the training set's time
range~\cite[\S~8.1]{arlot2010survey}.  In practice we find that while
Lasso and Ridge regressions both work well, it is preferable to use
Ridge regression as its implementation is often faster than Lasso on
the same data.

In \S\ref{sec:evaluation}, we compare the behaviour of the above
scoring functions, but we briefly explain their qualitative behaviour:
The univariate scoring mechanisms are cheaper to compute, but only
look at marginal dependencies between variables.  This can miss more
complex dependencies in data, some of which can only be ranked higher
when we look at joint and/or conditional dependencies.  Thus, the
joint mechanisms have {\em more statistical power} of detecting
complex dependencies between variables, but also run the risk of
over-fitting and producing more false-positives; Appendix~\ref{sec:r2}
gives more details about controlling false-positives.

%% \subsection{Interpreting ranking results}
%% The results of ranking can be used to infer the structure of the
%% causal BN.  Recall that each triple $(\X,\Y,\Z)$ in the result is
%% paired with its score.  When $\Z$ is empty, a high score means that
%% $\X\not\perp\Y$, which implies that $\X$ and $\Y$ are connected in the
%% causal BN.  When $\Z$ is not empty, a high score triples $(\X,\Y,\Z)$
%% means that $\X\not\perp\Y\mid\Z$, which can be used to infer the chain
%% structure, ruling out the fork structure rooted at $\Z$:
%% $\Y\leftarrow\Z\leftarrow\X$.  The main takeaway is that representing
%% queries using a triple $(\X,\Y,\Z)$ is sufficient to For more details
%% on deducing the possible structures, we refer the reader to the PC
%% algorithm~\cite{}.
 % 2pg
\section{Implementation}\label{sec:impl}
Our implementation had two primary requirements: It should be able to
integrate with a variety of data sources, such as OpenTSDB, Druid,
columnar data formats (e.g., parquet), and other data warehouses that
we might have in the future.  Second, it should be horizontally
scalable to test and score a large number of hypotheses.  Our target
scale was tens of thousands of hypotheses, with a response time to
generate a scoring report was in the order of a few minutes (for the
typical scale of hundreds of hypotheses) to an hour (for the largest
scale).

We implemented \sys using a combination of Apache
Spark~\cite{zaharia2016apache} and Python's scikit machine learning
library~\cite{scikit-learn}.  We used Apache Spark as a distributed
execution framework and to interface with external data sources such
as OpenTSDB, compressed parquet data files in our data warehouse, and
to plan and execute SQL queries using Spark
SQL~\cite{armbrust2015spark}.  We leveraged Python's scikit machine
learning library's optimised machine learning routines.  The user
interface is a web application that issues API calls to the backend
that specifies the input data, transformations, and display results to
the user.

In our use case, time series observations are taken every minute.
Most of our root cause analysis is done over 1--2 days of data, which
results in at most 1440--2880 data points per metric.  With $F$
features per family, the maximum dimension of the $\X_i$ feature
matrix is $2880 \times F$.  Realistically, we have seen (and tested)
scenarios up to $F \leq 80000$.  For $F$ in the order of tens of
thousands, the cost of {\em interpreting} the relevance of a group of
$F$ variables in a scenario already outweighs the benefit of doing a
joint analysis across all those variables.  For feature matrices in
this size range, a hypothesis can be scored easily on one machine;
thus, our unit of parallelisation is the hypothesis.  This avoids the
parallelisation cost and complexity of distributed machine learning
across multiple machines.  Thus, in our design each Spark executor
communicates to a local Python scikit kernel via IPC (we use Google's
{\tt gRPC}).

\subsection{Pipeline}\label{subsec:impl:pipeline}
The \sys pipeline can be broken down into three main stages.  In the
first stage, we implemented connectors in Java to interface with many
data sources to generate records, and User-Defined Functions (UDFs) in
Spark SQL to transform these records into a standardised Feature
Family Table (see Figure~\ref{fig:pipeline} for schema).  Thus, we
inherit Spark's support for joins and other statistical functions at
this stage.  In this first stage, users can write multiple Spark SQL
queries to integrate data from diverse sources, and we take the union
of the results from each query.  Then, we generate a Hypothesis Table
by taking a cross-product of the Feature Family Table and applying a
filter to select the target variable and the variables to condition.
In the final stage, we run a scoring function on the Hypothesis Table
to return the Top-K ($K=20$) results.  The Score Table also stores
plots for visualisation and debugging.  Appendix~\ref{sec:sql-queries}
lists the queries at various stages of the pipeline.

\subsection{Optimisations}
The declarative nature of the hypothesis query permits various
optimisations that can be deferred to the runtime system.  We describe
three such optimisations: Dense arrays, broadcast joins, and random
projections.

\smallsec{Dense arrays}: We converted the data in the Feature Family
Table into a {\tt numpy} array format stored in row-major order.  Most
of our time series observations are dense, but if data is sparse with
a small number of observations, we can also take advantage of various
sparse array formats that are compatible with the underlying machine
learning libraries.  This optimisation is significant: A na\"ive
implementation of our scorer on a single hypothesis triple in Spark
MLLib without array optimisations was at least 10x slower than the
optimised implementation in scikit libraries.

\smallsec{Broadcast join}: In most scenarios we have one target
variable $\Y$ and one set of auxiliary variables $\Z$ to condition
on. Hence, instead of a cross-product join on Feature Family Table, we
select $\Y$ and $\Z$ from the Feature Family table, and do a broadcast
join to materialise the Hypothesis Table.

\smallsec{Random projections}: To speed up multivariate hypothesis
testing (\S\ref{subsec:scalability}), we also use random projections
to reduce the dimensionality of features before doing penalised linear
regressions.  We sample a matrix $P_d$, a matrix of dimensions $T
\times d$, whose are drawn independently from a standard normal
distribution and project the data $(\X,\Y,\Z)$ into this a new space
$(P(\X),P(\Y),P(\Z))$ if the dimensionality of the matrix exceeds $d$;
that is,
\[
P(\X_{T\times n_x}) = {\begin{cases} \X & \text{if } n_x \leq d \\ \X P_d & \text{otherwise} \end{cases}}
\]
If we use random projections, we sample a new matrix every time we
project and take the average of three scores.  In practice, we find
there is little variance in these projections, so even one projection
is mostly sufficient for initial analysis.  Moreover, we prefer random
projection as it is simpler to implement, computationally more
efficient compared to dimensionality reduction techniques such as
Principal Component Analysis (PCA), with similar overall result
quality.  In some of our debugging sessions, we found that PCA
adversely impacted scoring.  This is because PCA reduces the feature
dimensionality by modeling the {\em normal} behaviour, and discards
the {\em anomalies} in the features that were needed to explain our
observations in the target variable.

\begin{table}[t]
\centering\small
\begin{tabularx}{\columnwidth}{l|p{0.65\columnwidth}}
  {\bf Component}          & {\bf Example causes} \\\hline
  Physical Infrastructure & Slow disks \\\hline
  Virtual Infrastructure  & NUMA issues, hypervisor network drops \\\hline
  Software Infrastructure & Kernel paging performance, Long JVM Garbage Collections \\\hline
  Services                & Slow dependent services \\\hline
  Input data              & Straggelers due to skew in data \\\hline
  Application code        & Memory leaks
\end{tabularx}
\caption{\sys hash helped us identify root-causes that belong to a
  diverse set of components.}\label{tab:diverse-issues}
\end{table}

\subsection{Asymptotic CPU cost}
For $T$ data points, and matrices of dimensions $T\times n_x$,
$T\times n_y$, and $T\times n_z$, denote the cost of doing a single
multivariate regression $\X \sim \Y$ as $C_{x,y} = \BigO(n_y \min(T
n_x^2, T^2 n_x))$.  Note that each joint/conditional regression runs
$k$ separate times for $k$-fold cross-validation, and does a
grid-search over $L$ values of the penalisation parameter for Ridge
regression.  Typically, $k$ and $L$ are small constants: $k=5$ and
$L=5$. Given these values, Table~\ref{tab:cost} lists the compute cost
for each scoring algorithm.

\begin{table}[h]
\centering\small
\begin{tabularx}{\columnwidth}{l|l}
  {\bf Method}          & {\bf Cost} \\\hline
  CorrMean, CorrMax     & $\BigO(n_x n_y T)$ \\
  Joint, Multivariate   & $\BigO(k L (C_{x,y} + C_{y,z} + C_{z,x}))$ \\
  Random Projection $d$ & $\BigO(k L T d (n_x + n_y + n_z + d))$
\end{tabularx}
\caption{The asymptotic CPU cost of scoring a hypothesis $(\X,\Y,\Z)$.
  As expected, the univariate method is the cheapest, and the joint
  and conditional methods are more expensive, with random projection
  into $d$ dimensions spanning the spectrum between the
  two.}\label{tab:cost}
\end{table}
 % 1pg
\section{Case Studies}\label{sec:case-study}
We now discuss a few case studies to illustrate how \sys helped us
diagnose the root-cause of undesirable performance behaviour.  In all
these examples, the setting is a more complex version of the example
in Figure~\ref{fig:simple-example}.  The main internal services
include tens of data processing and visualisation pipelines, operating
on over millions of events per second, writing data to the Hadoop
Distributed File System (HDFS).  Our key performance indicator is
overall runtime---the amount of time (in seconds) it takes to process
a minute's worth of input real time data to generate the final output.
This runtime is our target metric $\Y$ in all our case studies, and
the focus is on explaining runtimes that consistently average more
than a minute; these are problematic as it indicates that the system
is unable to keep up with the input rate.  Over the years, we found
that the root-cause for high runtimes were quite diverse spanning many
components as summarised in Table~\ref{tab:diverse-issues}.  Unless
otherwise mentioned, we start our analysis with feature families
obtained by grouping metrics by their name (and not any specific
key-value attribute).

% Controlled experiment: TCP retransmissions to understand how the
% system works.
\subsection{Controlled experiment: Injecting a fault into a live system}
In our first example, we discuss a scenario in which we injected a
fault into a live system.  Of all possible places we can introduce
faults, we chose the network as it affects almost every component
causing system-wide performance degradation.  In this sense, this
fault is an example of a hard case for our ranking as there could be a
lot of correlated effects.

\begin{figure}[t]
\centering
\includegraphics[width=0.35\textwidth]{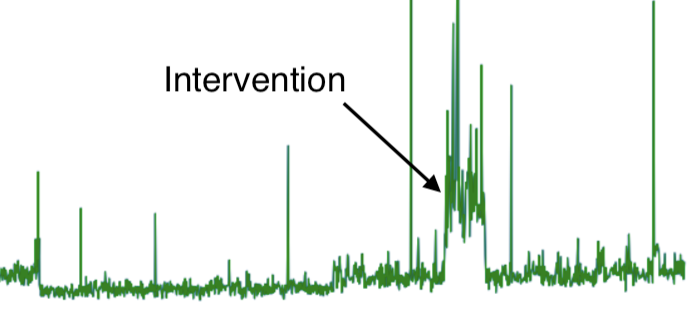}
\caption{A graph of pipeline runtime over time highlighting a period
  of high runtimes caused due to high packet retransmissions.}
\label{fig:intervention-retransmit}
\end{figure}

We injected packet drops at all datanodes by installing a Linux
firewall ({\small\tt iptables}) rule to drop 10\%\footnote{We chose 10\% as
  that was the smallest drop probability needed to cause a significant
  perceptual change in the observed runtime.} of all packets destined
to datanodes.  After a couple of minutes, we removed the firewall rule
and allowed the system to stabilise.
Figure~\ref{fig:intervention-retransmit} shows a screenshot of the
runtime time series, where the effect if dropping network packets is
clearly visible.

We ran \sys against all metrics in the system grouped by their name to
rank them based on the causal relevance to the observed performance
degradation (see Table~\ref{tab:intervention-retransmit} for the
ranking results).  The final results showed the following: (1) The
first set of metrics were the runtimes of a few other pipelines that
were ranked with high scores (about 0.7).  This was
expected, and we ignored these {\em effects} of the intervention.  (2)
The second set of metrics were the latencies of the above pipelines
whose runtimes were high.  Once again, these were expected since the
latency is a measure of the ``realtime-ness'' of the pipelines: the
difference between the current timestamp and the last timestamp
processed.

\begin{table}[t]
\centering\small
\begin{tabularx}{.46\textwidth}{l|p{.1\textwidth}|p{.25\textwidth}}
  {\bf Rank}     & {\bf Feature Family}    & {\bf Interpretation} \\\hline
  1--3, 5, 7
  & Runtime and latency of various pipelines
  & It took longer to save data.  Runtime is the sum of save times,
  so these dependencies are expected.\\\hline

  4
  & TCP Retransmit Count
  & Increased number of TCP retransmissions.
  \\\hline

  6
  & $75^{\rm th}$ percentile latency
  & Increase in database RPC latency.
  \\\hline

  8
  & Number of active jobs on the cluster
  & Increase in the number of active jobs scheduled on the cluster.
  \\\hline

  9
  & HDFS PacketAckRoundTrip time
  & Increase in the round-trip time for RPC acknowledgements between
  Datanodes.
\end{tabularx}
\caption{Global search across all metric families pinpointed to a
  network packet retransmission
  issue.}\label{tab:intervention-retransmit}
\end{table}

The third set of metrics were related to TCP retransmission counts
measured across all nodes in our cluster.  These counters, tracked by
the Linux kernel, measure the total number of packets that were
retransmitted by the TCP stack.  Packet drops induced by network
congestion, high bit error rates, and faulty cables are usually the
top causes when dealing on observing high packet retransmissions.  For
this scenario, these counters were clear evidence that pointed to a
network issue.

%% {\bf Remarks}: \sys also ranked a few other families of variables,
%% which were all effects correlated with the observed behaviour:
%% (1)~75th percentile RPC latency in the database system, (2)~the number
%% of queued jobs pending in the cluster, (3)~the round-trip time between
%% datanode RPCs, (4)~the number of active TCP connections, (5)~the
%% average finish times of all jobs in the cluster.  Although in this
%% extreme example we ran the search across all variables, the presence
%% of these additional variables correlated with the underlying cause
%% highlights the need for asking a specific hypothesis.
This example also showed us that although metrics in families 1--3, 5,
and 7 belonged to different groups by virtue of their names, they are
semantically similar and could be further grouped together in
subsequent user interactions.  The key takeaway is that \sys was able
to generate an explanation for the underlying behaviour (increased TCP
retransmissions).  In this case, the actual cause could be attributed
to packet drops that we injected, but as we shall see in the next
example, the real cause can be much more nuanced.

% Scale cluster performance improvement: conditioning on input sizes.
\subsection{The importance of conditioning: Disentangling multiple sources of variation}\label{subsec:conditioning}
Our next case study is a real issue we encountered in a production
cluster running at scale.  There was a performance regression compared
to an earlier version that was evident from high pipeline runtimes.
Although the two versions were not comparable (the newer version had
new functionality), it was important for us to understand what could
be done to improve performance.

% Overall look and behaviour.  What showed up?
We started by scoring all variables in the system against the target
pipeline runtime.  We found many explanations for variation.  At the
infrastructure level, CPU usage, network and disk IO activity, were
all ranked high.  At the pipeline service level, variations in task
runtimes, IO latencies, the amount of time spent in Java garbage
collection, all qualified as explanations for pipeline runtime to
various degrees of predictability.  Given the sheer scale of the
number of possible sources of variation, no single metric/feature
family served as a clear evidence for the degradation we observed.

% needed to condition on the input size.
To narrow down our search, we first noticed that it was reasonable to
expect high runtime at large scale.  Our load generator was using a
copy of actual production traffic that itself had stochastic
variation.  To separate out sources of variation into its constituent
parts, we {\em conditioned} the system state on the observed load size
prior to ranking.

% show ranking after conditioning.
The ranking had significantly changed after conditioning: The top
ranking families pointed to a network stack issue: metrics tracking
the number of retransmissions and the average network latency were at
the top, with a score of about 0.3.  However, unlike the
previous case-study, we did not know {\em why} there were packet
retransmissions but we were motivated to look for causes.
% show the retransmission behaviour.  count retransmissions per port.

% show how we followed up on this and saw that the buffers in VMs were
% filled up.  We increased the buffer size...
Since TCP packet retransmissions arise due to network packet drops, we
looked at packet drops at every layer in our network stack: at the
virtual machines (VM), the hypervisors, the network interface card on
the servers, and within the network.  Unfortunately, we could not
continue the analysis within \sys as we did not monitor these
counters.  We did not find drops within the network fabric, but one of
our engineers found that there were drops at the hypervisor's receive
queue because that the software network stack did not have enough CPU
cycles to deliver the packets to the VM.\footnote{We found that the
  {\tt time\_squeeze} counter in {\tt /proc/net/softnet\_stat} was
  continuously being incremented.}  Thus, we had a valid reason to
hypothesise that packet drops at the hypervisors were causing
variations in pipeline runtimes that were not already accounted for in
the size of the input.

\begin{figure}[t]
\centering
\includegraphics[width=0.35\textwidth]{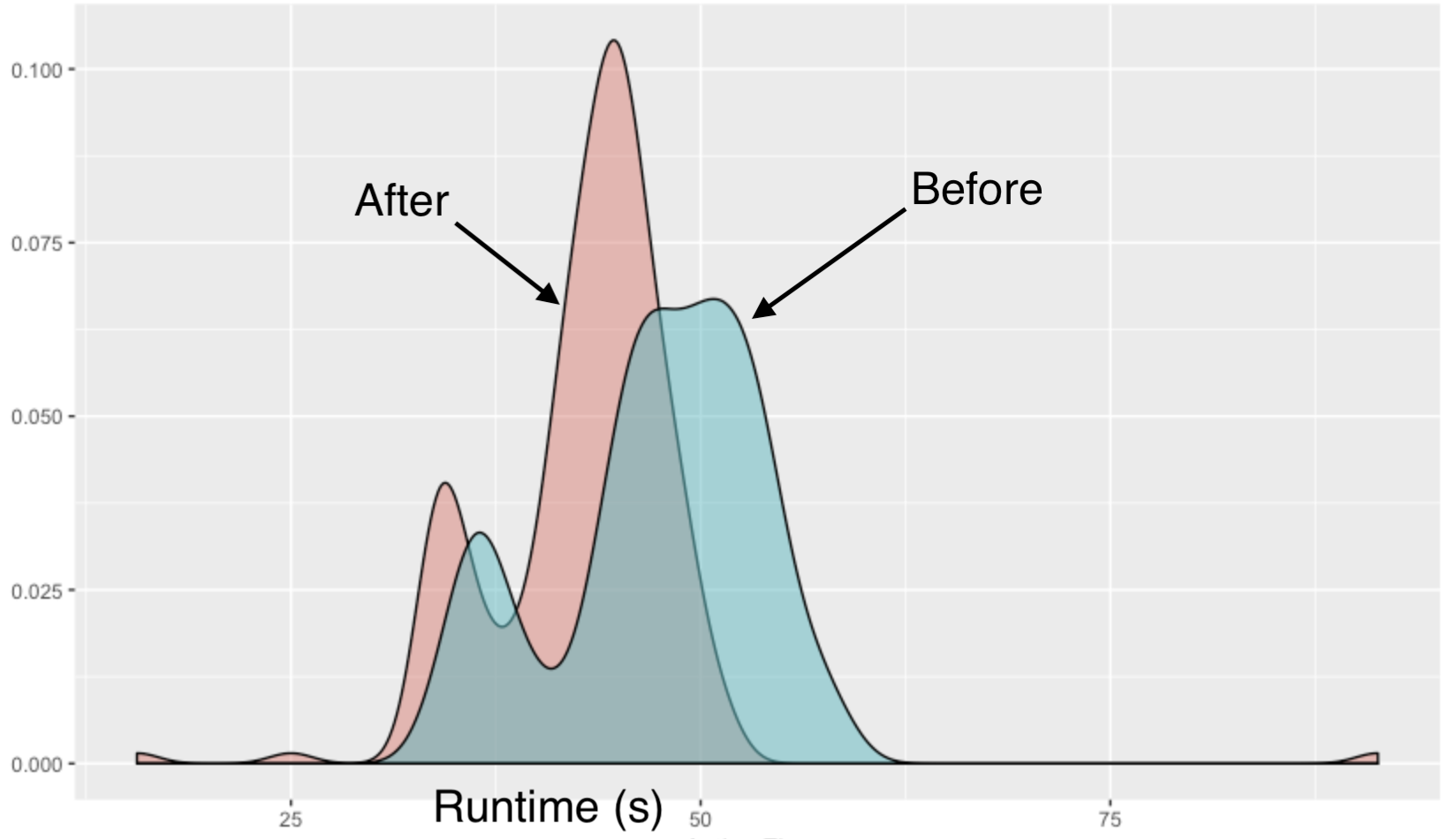}
\caption{Distributions of pipeline runtime for the same input data
  before and after the fix to reduce packet drops.  The bimodal nature
  of the graph is due to variations in input.}
\label{fig:perf-density}
\end{figure}

\smallsec{Experiment}: To establish a causal relationship, we
optimised our network stack to buffer more packets to reduce the
likelihood of packet drops.  After making this change on a live
system, we observed a 10\% reduction in the pipeline runtimes {\em
  across all pipelines}.  This experiment confirmed our hypothesis.
Figure~\ref{fig:perf-density} shows the distribution in runtime
before/after the change.  \sys's approach to {\em condition} on an
understood cause (input size) of variation in pipeline runtime helped
us debug a performance issue by focusing on alternate sources of
variation.  Although our monitored data was insufficient to
satisfactorily identify the root-cause (dropped packets at the
hypervisor), it helped us narrow it down sufficiently to come up with
a valid hypothesis that we could test.  By fixing the system, we
validated our hypothesis.  A second analysis after deploying the fix
showed that packet retransmissions was no longer the top ranking
feature; in fact the fix had eliminated packet drops.

%\begin{comment}
\ifarxiv
\subsection{Correlated with time: Periodic pipeline slowdown}\label{subsec:retention-spikes}
% Retention pipeline scanning HDFS every 15 minutes.
Our third case study is one in which there was a periodic spike in the
pipeline runtime, even when the cluster was running at less than 10\%
its peak load capacity.  On visual inspection, we saw that there was a
spike in the pipeline runtime from 10s to more than a minute every
(approximately) 15~minutes, and the spike lasted for about 5~minutes.
This abnormality was puzzling and pointed out to certain periodic activity
in the system.  We used \sys to find out the sources of variation and
found that metrics from the Namenode family were ranked high.  See
Table~\ref{tab:retention-spikes} for a summary of the ranking, and
Figure~\ref{fig:retention-spikes} for the behaviour.

\begin{table}[t]
\centering\small
\begin{tabularx}{.46\textwidth}{l|p{.1\textwidth}|p{.25\textwidth}}
  {\bf Rank}     & {\bf Feature Family}    & {\bf Interpretation} \\\hline

  1--4, 6--8
  & Runtime and latency of various pipelines
  & It took longer to save data.  Runtime is the sum of save times, so
  these variables are redundant.\\\hline

  5
  & Namenode metrics
  & Namenode service slowdown and degradation.
  \\\hline

  9
  & Detailed RPC-level metrics
  & Further evidence corroborating Namenode feature family at an RPC level.
  \\\hline

  27
  & JVM-level metrics
  & Increase in Datanode and Namenode waiting threads.
\end{tabularx}\caption{Global search across all metric families
  pinpointed to an issue at the Namenode.}\label{tab:retention-spikes}
\end{table}

When we narrowed our scope to Namenode metrics, we saw that there were
two classes of behaviour: positive and negative correlation with
respect to the pipeline runtime.  We observed that the Namenode's
average response latency was positively correlated with the pipeline
runtime (i.e., high response latency during high runtime intervals),
whereas Namenode Garbage Collection times were negatively correlated
to the runtime: i.e., smaller garbage collection when the pipeline
runtimes were high.  Thus, we ruled out garbage collection and tried
to investigate why the response latencies were high.

\begin{figure}[t]
\centering
\includegraphics[width=0.45\textwidth]{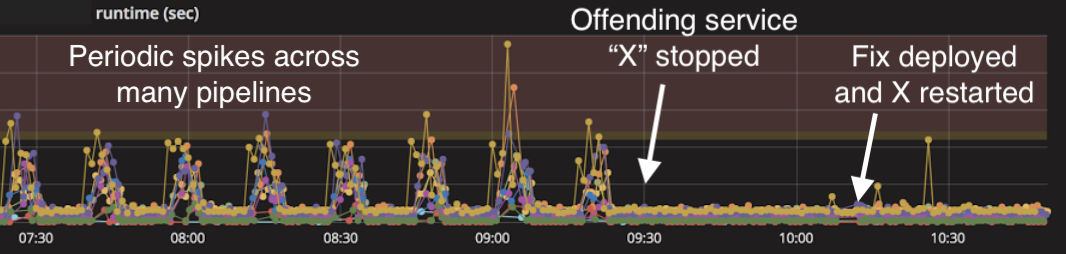}
\caption{Periodic spikes in the pipeline runtime (before 9:30)
  disappear after the offending service was fixed and restarted (at
  around 10:10).}
\label{fig:retention-spikes}
\end{figure}

A crucial piece of evidence was that the number of live processing
threads on the Namenode was also positively correlated with the
pipeline runtime.  Since the Namenode spawns a new thread for every
incoming RPC, we realised that a high request rate was causing the
Namenode to slow down.  We looked at the Namenode log messages and
observed a {\small\tt GetContentSummary} RPC call that was repeatedly
invoked; this prompted one engineer to suspect a particular service
that used this RPC call frequently.  When she looked at the code, she
found that the service made periodic calls to the Namenode with {\em
  exactly} the same frequency: once every 15~minutes.  These calls
were expensive because they were being used to scan the {\em entire
  filesystem}.

\smallsec{Experiment}: To test this hypothesis, the engineer quickly
pushed a fix that optimised the number of {\small\tt
  GetContentSummary} calls made by the service.  Within the next
15~minutes, we saw that the periodic spikes in latency had vanished,
and did not observe any more spikes.  This example shows how it is
important to reason about variations in metric behaviour with respect
to a model of how the system operates as the input load changes.  This
helped us eliminate Garbage Collection as a root-cause and dive deeper
into why there were more RPC calls.
%\end{comment}
\fi

%%%%%%%%%%%%%%%%%%%%%%%%%%%%%%
% Every week 4h high runtimes?
\subsection{Weekly spikes: Importance of time range}\label{subsec:weekly-spikes}
Our final example illustrates another example of pipeline runtime that
was correlated with time: occasionally, all pipelines would run slow.
We observed no changes in input sizes (a handful of metrics that we
monitor along with the runtime) that could have explained this
behaviour, so we used \sys to dive deeper.  The top five feature
families are shown in Table~\ref{tab:weekly-spikes}.  We dismissed the
first two feature families as irrelevant to the analysis because the
variables were effects, which we wanted to explain in the first place.
The third and fourth variables were interesting.  When we reran the
search to rank variables restricting the search space to only load and
disk utilisation, we noticed that the hosts that ran our datanodes
explained the increase in runtimes with high score.  However, \sys did
not have access to per-process disk usage, so we resorted to
monitoring the servers manually to catch the offender.  Unfortunately,
the issue never resurfaced in a reasonable amount of time.

\begin{table}[t]
\centering\small
\begin{tabularx}{.5\textwidth}{l|p{.1\textwidth}|p{.3\textwidth}}
  {\bf Rank}     & {\bf Feature Family}    & {\bf Interpretation} \\\hline
  1              & Pipeline data save time & It took longer to save data.
  Runtime is the sum of save times,
  so this variable is redundant. \\\hline

  2              & Indexing component runtime & It took longer to index data.
  The effect is not localised, but shared across all components.
  \\\hline

  3              & Increase in load average   & More than usual Linux processes
  were waiting in the scheduler run queue.
  \\\hline

  4              & Increase in disk utilisation & High disk IO coinciding with spikes.
  \\\hline

  5, 6           & Latency, derived from families 1 and 2  & Increase in runtime increases
  latency, so this is expected.
  \\\hline

  7              & RAID monitoring data            & Spikes in temperature
  recorded by the RAID controller.
\end{tabularx}\caption{Global search across all metric families
  pinpointed to a disk IO issue.}\label{tab:weekly-spikes}
\end{table}

\begin{figure}[h]
\centering
\includegraphics[width=0.45\textwidth]{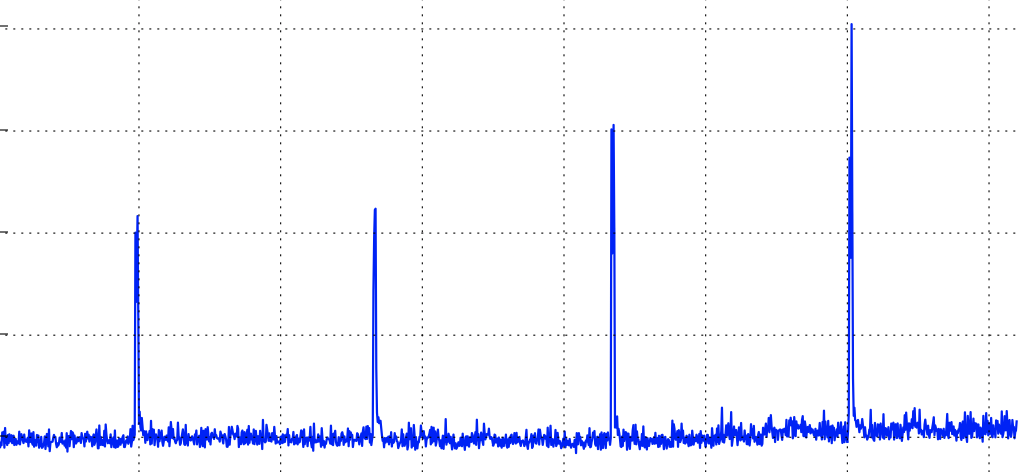}
\caption{Weekly spikes in pipeline runtime when viewing across a time
  range of a month.}
\label{fig:weekly-spikes}
\end{figure}

However, these issues occurred sporadically across many of our
clusters.  When we looked at time ranges of over a month, we noticed a
regularity in the spikes: they had a period of 1~week, and it lasted
for about 4~hours (see Figure~\ref{fig:weekly-spikes}).  Since we
could not account the disk usage to any specific Linux process, we
suspected that there was an infrastructure issue.  We asked the
infrastructure team what could potentially be happening every week,
and one engineer had a compelling hypothesis: Our disk hardware was
backed by hardware redundancy (RAID).  There is a periodic disk
consistency check that the RAID controller performs every 168~hours
(1~week!)~\cite{raid-weekly}.  This consistency check consumes disk
bandwidth, which could potentially affect IO bandwidth that is
actually available to the server.  The RAID controller also provided
knobs to tune the maximum disk capacity that is used for these
consistency checks.  By default, it was set to 20\% of the disk IO
capacity.

\begin{figure}[t]
\centering
\includegraphics[width=0.3\textwidth]{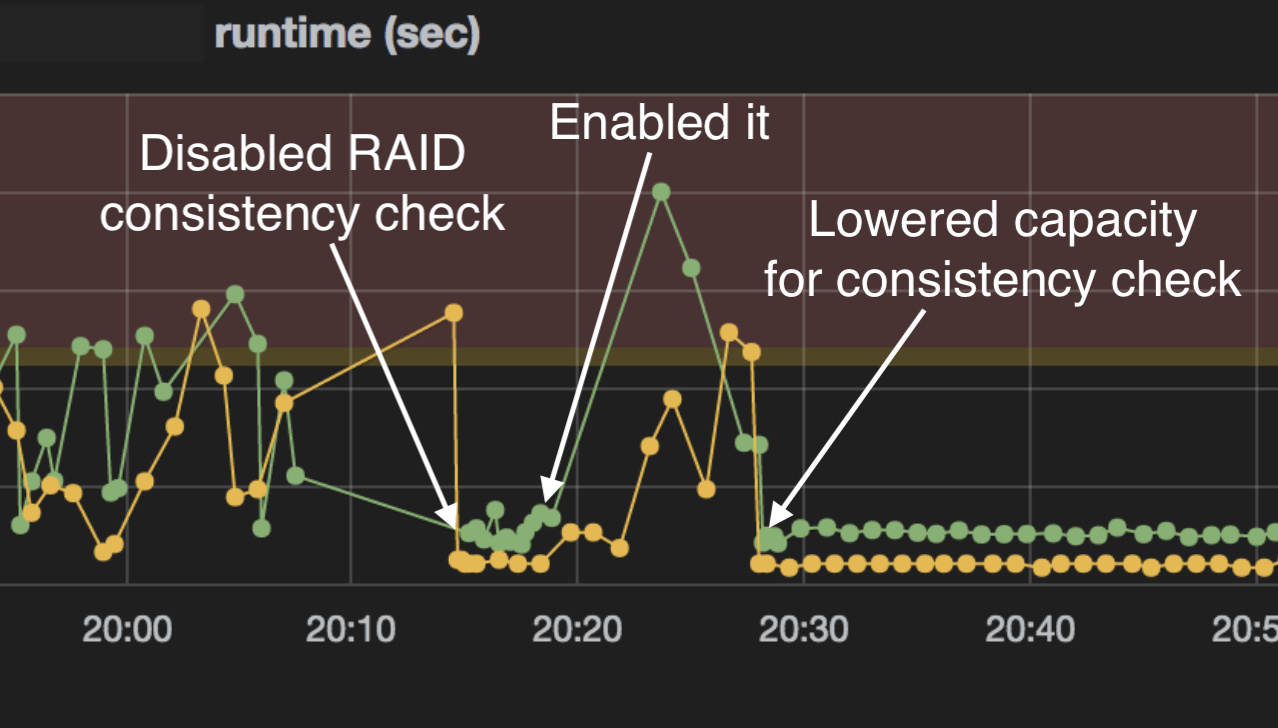}
\caption{Results of an intervention on a live system to test the
  hypothesis that a specific RAID controller setting was causing
  periodic performance slowdown.}
\label{fig:raid-experiment}
\end{figure}

\smallsec{Experiment}: Once we had a hypothesis to check, we waited
for the next predicted occurrence of this phenomenon on a cluster.  We
were able to perform two controlled experiments: (1)~disable the
consistency check, and (2)~reduce disk IO capacity that the
consistency checks use to 5\%.  Figure~\ref{fig:raid-experiment} shows
the results of the intervention.  From 2000~hrs to 2015~hrs, the
cluster was running with the default configuration, where the runtimes
showed instabilities.  From 2015~hrs to 2020~hrs we disabled the
consistency check, before re-enabling it at 2020~hrs.  Finally, at
2025~hrs, we reduced the maximum capacity for consistency checks to
5\%.  This experiment confirmed the engineer's hypothesis, and a fix
for this issue went immediately into our product.
 % 3pg

\begin{table*}[t]
\centering
\begin{tabular}{r|r|r|r|r|r|r|r}
  {\bf Scenario \#}
  & {\bf \# Families}
  & {\bf \# Features }
    & {\bf CorrMean}
    & {\bf CorrMax}
    & $\pmb{L_2}$
    & $\pmb{L_2-P50}$
    & $\pmb{L_2-P500}$  \\\hline
1 & 816 & 130259 & 0.167 & 1.000 & 0.143 & 1.000 & 0.333 \\\hline
2 & 2337 & 158253& 0.143 & 0.071 & - & 0.077 & - \\\hline
3 & 902 & 61229 & 1.000 & 1.000 & 0.200 & 1.000 & 1.000 \\\hline
4 & 2156 & 141082& - & - & 0.333 & 0.167 & 0.333 \\\hline
5 & 800 & 63797 & - & 1.000 & 0.100 & 1.000 & 0.077 \\\hline
6 & 436 & 29689 & - & - & 0.333 & 0.167 & 0.500 \\\hline
7 & 751 & 61231 & - & 0.111 & 1.000 & - & 0.200 \\\hline
8 & 603 & 100486 & - & 1.000 & 0.250 & 1.000 & 1.000 \\\hline
9 & 622 & 51230 & 0.050 & 0.053 & 0.500 & 0.062 & 0.250 \\\hline
10 & 601 & 71227  & - & 0.500 & 1.000 & 0.333 & 0.250 \\\hline
11 & 509 & 27902  & 0.333 & 0.083 & - & - & - \\\hline\hline
\multicolumn{3}{r|}{\bf Summary}
    & {\bf CorrMean}
    & {\bf CorrMax}
    & $\pmb{L_2}$
    & $\pmb{L_2-P50}$
    & $\pmb{L_2-P500}$  \\\hline\hline
\multicolumn{3}{r|}{Harmonic mean (discounted gain)} & 0.002 & 0.004 & {\cf 0.009} & {\cf 0.009} & {\cf 0.009} \\\hline
\multicolumn{3}{r|}{Average (discounted gain)} & 0.154 & {\cf 0.438} & 0.351 & {\cf 0.437} & 0.359 \\\hline
\multicolumn{3}{r|}{Stdev of average discounted gain} & 0.300 & 0.465 & 0.353 & 0.456 & 0.350 \\\hline
\multicolumn{3}{r|}{Perfect score / success (\%) top-1}  & 7  & {\cf 23} & 15 & {\cf 23} & 15 \\\hline
\multicolumn{3}{r|}{Success (\%) top-5}  & 19 & 46 & 64 & 46 & {\cf 73} \\\hline
\multicolumn{3}{r|}{Success (\%) top-10} & 37 & 55 & {\cf 82} & 64 & 73 \\\hline
\multicolumn{3}{r|}{Success (\%) top-20} & 46 & {\cf 82} & {\cf 82} & {\cf 82} & {\cf 82} \\\hline
\end{tabular}
\caption{A summary of the sizes of input datasets, and performance of
  various scoring methods.  The feature family grouping is by the name
  of the metric.  The mean number of features per feature family in a
  scenario varies between ~50--180, and the maximum is between
  ~2000--75000.  For each scenario, we compute the discounted gain, a
  measure of ranking accuracy.  The summary shows that both CorrMax
  and $\pmb{L_2-P50}$ work quite well, with $\pmb{L_2-P50}$ being a
  superior method that has power to detect joint effects just like
  $\pmb{L_2}$.  The failures are marked with a hyphen; we use a small
  score of $\pmb{0.001}$ when including failures for computing the
  harmonic mean summary.  Note that in all cases given the large
  number of features, a random ranking results in a low score (much
  worse than CorrMean).  The boldface highlighted numbers are the best
  overall results.}\label{tab:eval}
\end{table*}

\section{Evaluation}\label{sec:evaluation}
We now focus on more quantitative evaluation of various aspects of
\sys.  We find that the declarative aspect of \sys simplifies
generating tens of thousands of hypotheses at scale with a handful of
queries.  Moreover, we find that no single scorer dominates the other:
each algorithm has its strengths and weaknesses:
\begin{itemize}[noitemsep,leftmargin=1em,topsep=1pt]
  \item Univariate scoring has low false positives, but also has low
    statistical power; i.e., fails to detect explanations for
    phenomena that involve multiple variables jointly.
  \item Joint scoring using penalised regression is slower, and the
    ranking is biased towards feature families that have a large
    number of variables, but has more power than univariate scoring.
  \item Random projection strikes a tradeoff between speed and
    accuracy and can rank causes higher than other joint methods.
\end{itemize}

We run our tests on a small distributed environment that has about 8
machines each with 256GB memory and 20 CPU cores: the Spark executors
are constrained to 16GB heap, and the remaining system memory can be
used by the Python kernels for training and inference.  These machines
are shared with other data processing pipelines in our product, but
their load is relatively low.

\subsection{Scorers}
We took data from 11 additional root-cause incidents in our
environment and compared various scoring methods on their efficacy.
None of these incidents needed conditioning.  Table~\ref{tab:eval}
shows some summary statistics about each incident.  We compare the
following five scoring methods:
\begin{itemize}[noitemsep,leftmargin=1em,topsep=0pt]
\item CorrMean: mean absolute pairwise correlation,
\item CorrMax: max absolute pairwise correlation,
\item $L_2$: joint ridge regression scoring,
\item $L_2-P50$: joint ridge regression after projecting to (at most)
  50 dimensional space,
\item $L_2-P500$: joint ridge regression after projecting to (at most)
  500 dimensional space ($L_2-P500$).
\end{itemize}

\begin{figure*}[t]
\centering
\includegraphics[width=\columnwidth]{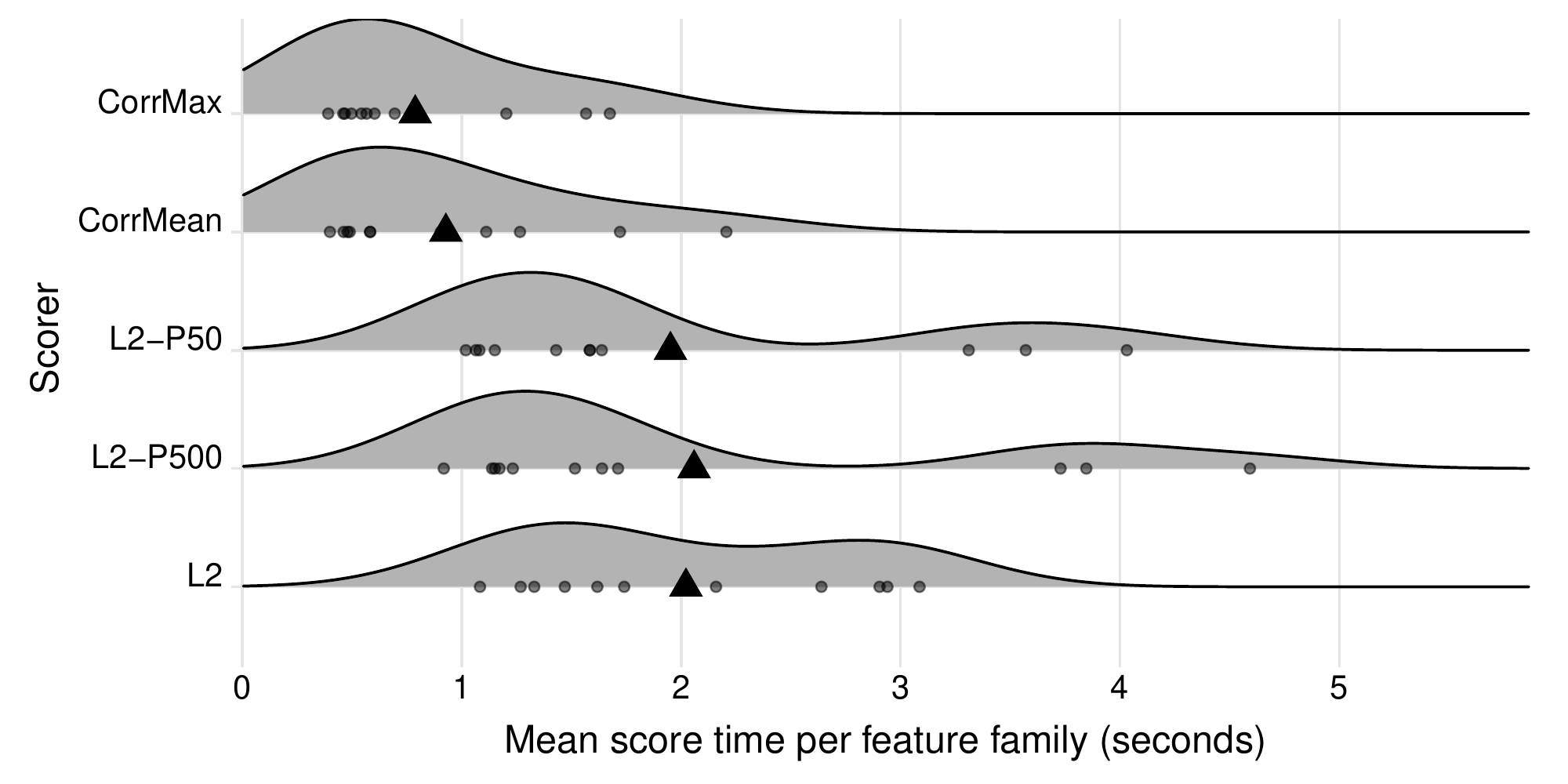}
\includegraphics[width=\columnwidth]{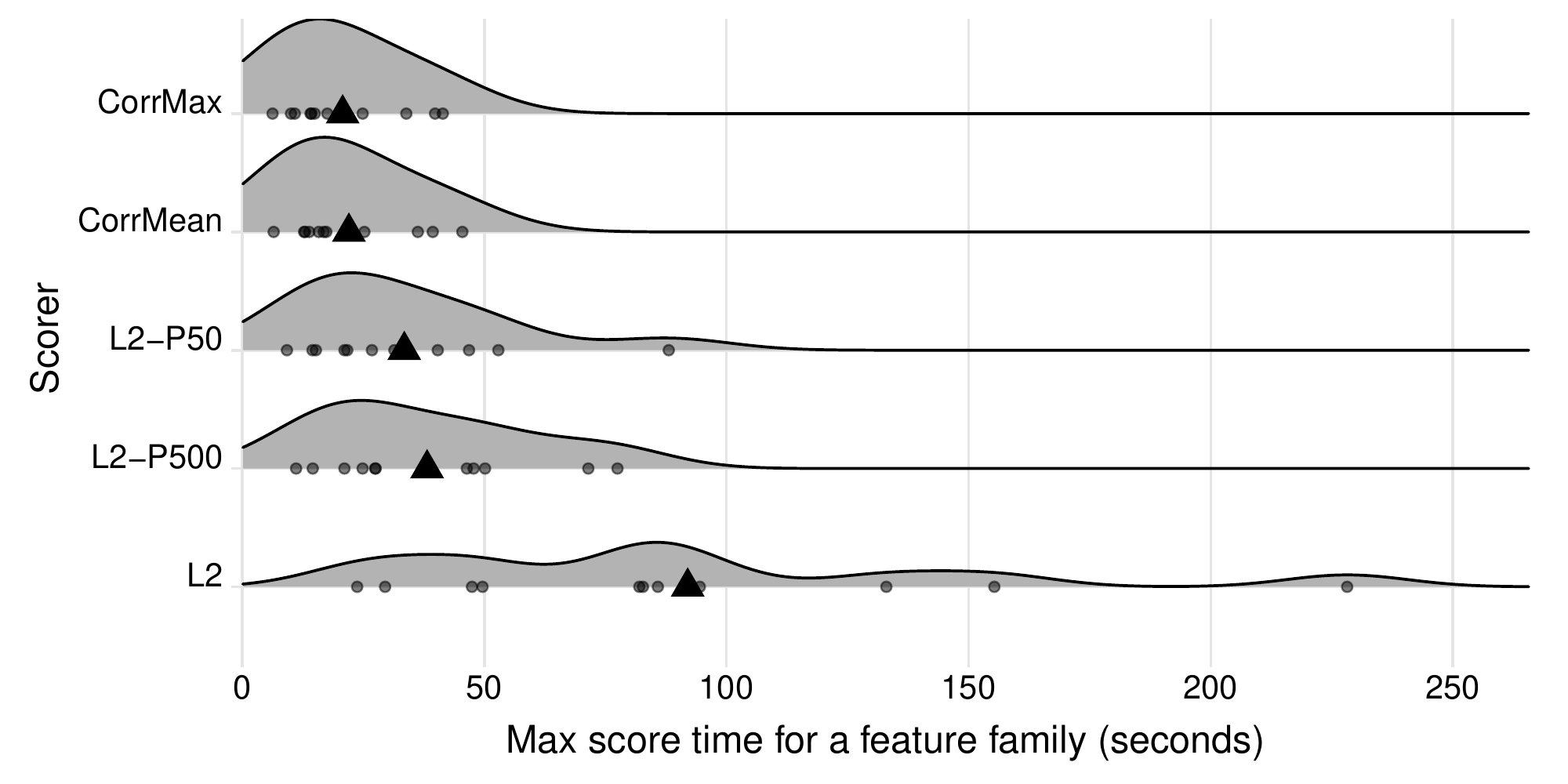}
\caption{Density plot of runtimes of all scenarios, normalised to mean
  (top) and max (bottom) score time per feature family (regardless of
  the number of features) for various scoring techniques.  All
  multivariate techniques use $k=5$-fold cross-validation, a grid
  search over 3~values of the ridge regression penalty hyper-parameter.
  Random projection returns the average score of 3~random samples of
  the projection matrix.  The data points are marked with $\bullet$,
  and the mean of each distribution is marked with $\blacktriangle$.}
\label{fig:eval:runtime}
\end{figure*}

%% \begin{table}[t]
%% \centering\small
%% \begin{tabular}{l|l|l}
%%   {\bf Scenario} & {\bf \# Families}    & {\bf \# Features} \\\hline
%% S1 & 816 & 130259 \\\hline
%% S2 & 2337 & 158253 \\\hline
%% S3 & 902 & 61229 \\\hline
%% S4 & 2156 & 141082 \\\hline
%% S5 & 800 & 63797 \\\hline
%% S6 & 436 & 29689 \\\hline
%% S7 & 751 & 61231 \\\hline
%% S8 & 603 & 100486 \\\hline
%% S9 & 622 & 51230 \\\hline
%% S10 & 601 & 71227 \\\hline
%% S11 & 509 & 27902 \\\hline
%% \end{tabular}
%% \caption{Sizes of various scenarios; each feature family is grouped by
%%   the name of the metric.  The mean number of features per family
%%   varies between ~50--180, and the maximum between
%%   ~2000--75000.}\label{tab:eval:dataset}
%% \end{table}
We manually labelled only the top-20 results in each scenario as
either a cause, an effect, or irrelevant (happens only for scores).
The scores in top-20 were large enough that no variables were marked
irrelevant. To compare methods, we look at the following metrics for a
single scenario:
\begin{itemize}[noitemsep,leftmargin=1em,topsep=0pt]
  \item {\bf Ranking accuracy}: If $r$ is the rank of the first cause,
    define the accuracy to be $1/r$.  This measures the discounted
    ranking gain~\cite{jarvelin2002cumulated,wang2013theoretical},
    with a binary relevance of 0 for effect, 1 for cause, and a
    Zipfian discount factor of $1/r$ (cutoff of top-20).  We also
    report the arithmetic and harmonic mean of accuracy across
    scenarios.
  \item {\bf Success rate} (in top-$k$): Define precision $p$ for a
    single scenario as 1 if there is a cause in the top $k$ results, 0
    otherwise.  We also report average success rate (across scenarios)
    of the top-$k$ ranking for various $k$.
\end{itemize}

Table~\ref{tab:eval} shows the results.  The experiments reveal a few
insights, which we discuss below.  First, univariate scoring methods
complement the joint scoring methods that are not robust to feature
families with a large number of features.  Univariate methods shine
well if the cause itself is univariate.  However, multivariate methods
outperform univariate methods if, by definition, there are multiple
features that jointly explain a phenomenon (e.g.,
\S\ref{subsec:weekly-spikes}).  On further inspection, we found that
the true causes did have a {\em non-zero} score in the multivariate
scorer, but they were ranked lower and hence did not appear in top-20.
Second, random projection serves as a good balance of tradeoff between
univariate methods and multivariate joint methods.  We observed a
similar behaviour for discounted cumulative ranking gain with a
discount factor of $1/\log(1+r)$ instead of $1/r$.

\smallsec{Takeaway}: The complementary strengths of the two methods
highlight how the user can choose the inexpensive univariate scoring
if they have reasons to believe that a single univariate variable
might be the cause, or the more expensive multivariate scoring if they
are unsure.  This tradeoff further demonstrates how declarative
queries can be exploited to defer such decisions to the runtime
system.  We are working on techniques to automatically select the
appropriate method without user intervention.

\subsection{Scalability}\label{subsec:scalability}
Since \sys supports adhoc queries for generating hypotheses from many
data sources, the end-to-end runtime depends on the query and size of
the input dataset, the number of scored hypotheses/feature families,
and the number of metrics per hypothesis.  We found that the scoring
time is predominantly determined by the number of hypotheses, and
hence normalise the runtime for the 11~scenarios listed above per
feature family.  Figure~\ref{fig:eval:runtime} shows the scatter plot
of scenario runtimes for the five different scoring algorithms.
Despite multivariate techniques being computationally expensive, the
actual runtimes are within a 2--3x of the simpler scorer (on average),
and within 1.5x (for max).  Note that this cost {\em includes} the
data serialisation cost of communicating the input matrix and the
score result between the Java process and the Python process, which
likely adds a significant cost to computing the scores.  On further
instrumentation, we find that serialisation accounts on average about
25\% of the total score time per feature family for the univariate
scorers, and only about 5\% for the multivariate joint scorers.
 % 1pg
\section{Related Work}\label{sec:relwork}
\sys builds on top of fundamental techniques and insights from a large
body of work that on troubleshooting systems from data.  To our
knowledge, \sys is the first system to conduct and report analysis at
a large scale.

\smallsec{Theoretical work}: Pearl's work on using graphical models as
a principled framework for causal inference~\cite{pearl2009causality}
is foundation for our work.  Other algorithms for causal discovery
such as PC/SGS~\cite[Sec.~5.4.1]{spirtes2000causation} algorithm,
LiNGAM~\cite{shimizu2006linear} all use pairwise conditional
independence tests to discover the full causal structure; we showed
how key ideas from the above works can be improved by also considering
a joint set of variables.  As we saw in \S\ref{sec:approach},
root-cause analysis in a practical setting rarely requires the full
causal structure of the data generating process.  Moreover, we
simplified identifying a cause/effect by taking advantage of metadata
that is readily available, and by allowing the user to query for
summaries at a desired granularity that mirrors the system structure.

% ML approaches: L1 regression sparsity, etc.
% explainable ML?
\begin{comment}
\smallsec{Explainable models}: \sys differs from the machine learning
approaches designed with a goal of producing sparse, interpretable
models.  For instance, the LASSO aims to select a small subset of
predictors that best explain a target variable (in the least squares
sense).  However, the approach of using all features in LASSO can be
problematic because LASSO is known to arbitrarily select one among
several correlated predictors.  \sys takes an alternate approach by
placing the user in the loop to guide the search for hypotheses.
\end{comment}
% Other systems to search for similar time series.
%% Similarity based approaches that search for time series similar to a
%% target query~\cite{ding2008querying}.  The Pearson correlation
%% coefficient can be thought of a simple approach that falls in this
%% category.  Pairwise relations are but a step towards quantifying the
%% association between a pair of variables; we emphasised the need for a
%% {\em principled} approach to effectively summarise such associations
%% between a feature family and a target, opening up a large class of
%% techniques that can be selected at runtime.

\smallsec{Systems}: \sys is an example of a tool for Exploratory Data
Analysis~\cite{tukey1977exploratory}, and one recent work that shares
our philosophy is MacroBase~\cite{bailis2017macrobase}.  MacroBase
makes a case for prioritising attention to cope with the volume of
data that we generate, and prioritising rapid interaction with the
user to enable better decision making.  \sys can be thought of as a
specific implementation of the key ideas in MacroBase for root-cause
analysis, with additional techniques (conditioning and pseudo-causes)
to further prioritise attention to specific variations in the data.

% Python/R formulae
The declarative way of specifying hypotheses in \sys is largely
inspired by the {\em formula} syntax in the R language for statistical
computing~\cite{r-formula-1,r-formula-2}.  In a typical R workflow for
model fitting, a user prepares her data into a tabular data-frame
object, where the rows are observations and the columns are various
features.  The formula syntax is a compact way to specify the
hypothesis in a declarative way: the user can specify conditioning,
the target features, interactions/transformations of those features,
lagged variables for time series~\cite{r-formula-3}, and
hierarchical/nested models.  However, this formula still refers to
{\em one} model/hypothesis. \sys takes this idea further to use SQL to
generate the candidate models.

\smallsec{Practical experience}: Prior tools designed for a specific
use-case rely on labelled data (e.g., \cite{deb2017aesop} for network
operators), which we did not have when encountering failure modes for
the first time.  \sys also employs hierarchies to scale understanding
(similar to~\cite{nair2015learning}); however, we demonstrated the
need for conditioning to filter out uninteresting events.  Early
work~\cite{cohen2004correlating} proposed using a tree-augmented
Bayesian Network as a building block for automated system diagnosis.
Our experience is that a hierarchical model of system behaviour needs
to be continuously updated to reflect the reality.  \sys is
particularly useful in bootstrapping new models when the old model
does not reflect reality.

Another line of work on time series
data~\cite{chen2008exploiting,tao2014metric,cheng2016ranking} has
focused primarily on {\em detecting} anomalies, by looking for
vanishing (weakening) correlations among variables (when an anomaly
occurs) \cite{chen2008exploiting}. Subsequent work uses similar
techniques to both detect and rank possible causes based on timings of
change propagation or other features of time series'
interactions~\cite{cheng2016ranking,tao2014metric}. In our use cases,
we have often found a diversity of causes, and existing correlations
among variables do not weaken sufficiently during a period of
interest.  Moreover, our work also shows the importance of human in
the loop to discern the likely causes from what is shown, and by
further interaction and conditioning as necessary.
 % 1pg
\section{Conclusions}
When we started this work, our goal was to build a data-driven
root-cause analysis tool to speed up troubleshooting to harden our
product.  Our experience in building it taught us that the fewer
assumptions we make, the better the tool generalises.  Over the last
four years, the diversity of troubleshooting scenarios taught us that
it is hard to completely automate root-cause analysis without humans
in the loop.  The results from \sys helped us not only identify
issues, but also fix them.  We found that the time series metadata
(names and tags) has a rich hierarchical structure that can be
effectively utilised to group variables into human-relatable entities,
which in practice we found to be sufficient for explainability.  We
are continuing to develop \sys and incorporate other sources of data,
particularly text time series (log messages), and also improving the
ranking using results multiple queries.

%\balance

%ACKNOWLEDGMENTS are optional
%% \section*{Acknowledgements}
%% We thank the entire team at Tetration Analytics who play a critical
%% role in building, managing, and deploying the infrastructure and
%% collecting the data necessary for building \sys.  Vimalkumar would
%% also like to thank Arun Kumar for his comments on an early draft that
%% helped improve the presentation.

% The following two commands are all you need in the
% initial runs of your .tex file to
% produce the bibliography for the citations in your paper.
\bibliographystyle{abbrv}
\bibliography{explainit}

\begin{thebibliography}{10}

\bibitem{r-formula-3}
{Dynamic Linear Models and Time-Series Regression}.
\newblock
  \url{http://math.furman.edu/~dcs/courses/math47/R/library/dynlm/html/dynlm.html}.

\bibitem{explainit-extended}
{ExplainIt! -- A declarative root-cause analysis engine for time series data
  (extended version)}.
\newblock \url{https://arxiv.org/abs/1903.08132}.

\bibitem{fred}
{FRED: Economic Research Data}.
\newblock \url{https://fred.stlouisfed.org/}.

\bibitem{raid-weekly}
{LSi Megaraid Patrol Read and Consistency Check schedule recommendations}.
\newblock
  \url{https://community.spiceworks.com/topic/1648419-lsi-megaraid-patrol-read-and-consistency-check-schedule-recommendations}.

\bibitem{wikiconditional}
{Multivariate normal distribution: Conditional distributions}.
\newblock
  \url{https://en.wikipedia.org/wiki/Multivariate_normal_distribution#Conditional_distributions}.

\bibitem{opentsdb}
{OpenTSDB: Open Time Series Database}.
\newblock \url{http://opentsdb.net}.

\bibitem{nasa}
{Prognostic Tools for Complex Dynamical Systems}.
\newblock
  \url{https://www.nasa.gov/centers/ames/research/technology-onepagers/prognostic-tools.html}.

\bibitem{r-formula-1}
{R: Model Formulae}.
\newblock
  \url{https://www.rdocumentation.org/packages/stats/versions/3.5.1/topics/formula}.

\bibitem{r-formula-2}
{Statistical formula notation in R}.
\newblock
  \url{http://faculty.chicagobooth.edu/richard.hahn/teaching/formulanotation.pdf}.

\bibitem{vmware-wavefront}
{vmWare WaveFront}.
\newblock \url{https://www.wavefront.com/user-experience/}.

\bibitem{stackexchange-r2}
{What is the distribution of $r^2$ in OLS?}
\newblock \url{https://stats.stackexchange.com/a/130082}.

\bibitem{arlot2010survey}
S.~Arlot, A.~Celisse, et~al.
\newblock {A survey of cross-validation procedures for model selection}.
\newblock {\em Statistics surveys}, 2010.

\bibitem{armbrust2015spark}
M.~Armbrust, R.~S. Xin, C.~Lian, Y.~Huai, D.~Liu, J.~K. Bradley, X.~Meng,
  T.~Kaftan, M.~J. Franklin, A.~Ghodsi, et~al.
\newblock {Spark sql: Relational data processing in spark}.
\newblock {\em SIGMOD}, 2015.

\bibitem{sep-physics-Rpcc}
F.~Arntzenius.
\newblock {Reichenbach's Common Cause Principle}.
\newblock {\em The Stanford Encyclopedia of Philosophy}, 2010.

\bibitem{bailis2017macrobase}
P.~Bailis, E.~Gan, S.~Madden, D.~Narayanan, K.~Rong, and S.~Suri.
\newblock {Macrobase: Prioritizing attention in fast data}.
\newblock {\em SIGMOD}, 2017.

\bibitem{barten1962note}
A.~Barten.
\newblock {Note on unbiased estimation of the squared multiple correlation
  coefficient}.
\newblock {\em Statistica Neerlandica}, 1962.

\bibitem{benjamini1995controlling}
Y.~Benjamini and Y.~Hochberg.
\newblock {Controlling the false discovery rate: a practical and powerful
  approach to multiple testing}.
\newblock {\em Journal of the royal statistical society. Series B
  (Methodological)}, 1995.

\bibitem{chen2008exploiting}
H.~Chen, H.~Cheng, G.~Jiang, and K.~Yoshihira.
\newblock {Exploiting local and global invariants for the management of large
  scale information systems}.
\newblock {\em ICDM}, 2008.

\bibitem{cheng2016ranking}
W.~Cheng, K.~Zhang, H.~Chen, G.~Jiang, Z.~Chen, and W.~Wang.
\newblock {Ranking causal anomalies via temporal and dynamical analysis on
  vanishing correlations}.
\newblock {\em SIGKDD}, 2016.

\bibitem{cohen2004correlating}
I.~Cohen, J.~S. Chase, M.~Goldszmidt, T.~Kelly, and J.~Symons.
\newblock {Correlating Instrumentation Data to System States: A Building Block
  for Automated Diagnosis and Control.}
\newblock {\em OSDI}, 2004.

\bibitem{cramer1987mean}
J.~S. Cramer.
\newblock {Mean and variance of R2 in small and moderate samples}.
\newblock {\em Journal of econometrics}, 1987.

\bibitem{deb2017aesop}
S.~Deb, Z.~Ge, S.~Isukapalli, S.~Puthenpura, S.~Venkataraman, H.~Yan, and
  J.~Yates.
\newblock {AESOP: Automatic Policy Learning for Predicting and Mitigating
  Network Service Impairments}.
\newblock {\em SIGKDD}, 2017.

\bibitem{eaton1983multivariate}
M.~L. Eaton.
\newblock {Multivariate statistics: a vector space approach.}
\newblock {\em Wiley}, 1983.

\bibitem{jarvelin2002cumulated}
K.~J{\"a}rvelin and J.~Kek{\"a}l{\"a}inen.
\newblock {Cumulated gain-based evaluation of IR techniques}.
\newblock {\em TOIS}, 2002.

\bibitem{explainit-causalml2018}
V.~Jeyakumar, O.~Madani, A.~Parandeh, A.~Kulshreshtha, W.~Zeng, and N.~Yadav.
\newblock {ExplainIt!: Experience from building a practical root-cause analysis
  engine for large computer systems}.
\newblock {\em CausalML Workshop, ICML}, 2018.

\bibitem{koerts1969theory}
J.~Koerts and A.~P.~J. Abrahamse.
\newblock {On the theory and application of the general linear model}.
\newblock 1969.

\bibitem{nair2015learning}
V.~Nair, A.~Raul, S.~Khanduja, V.~Bahirwani, Q.~Shao, S.~Sellamanickam,
  S.~Keerthi, S.~Herbert, and S.~Dhulipalla.
\newblock {Learning a hierarchical monitoring system for detecting and
  diagnosing service issues}.
\newblock {\em SIGKDD}, 2015.

\bibitem{olejnik2000using}
S.~Olejnik, J.~Mills, and H.~Keselman.
\newblock {Using Wherry's adjusted R 2 and Mallow's Cp for model selection from
  all possible regressions}.
\newblock {\em The Journal of experimental education}, 2000.

\bibitem{pearl2009causality}
J.~Pearl.
\newblock {Causality}.
\newblock 2009.

\bibitem{scikit-learn}
F.~Pedregosa, G.~Varoquaux, A.~Gramfort, V.~Michel, B.~Thirion, O.~Grisel,
  M.~Blondel, P.~Prettenhofer, R.~Weiss, V.~Dubourg, J.~Vanderplas, A.~Passos,
  D.~Cournapeau, M.~Brucher, M.~Perrot, and E.~Duchesnay.
\newblock {Scikit-learn: Machine Learning in {P}ython}.
\newblock {\em JMLR}, 2011.

\bibitem{pelkonen2015gorilla}
T.~Pelkonen, S.~Franklin, J.~Teller, P.~Cavallaro, Q.~Huang, J.~Meza, and
  K.~Veeraraghavan.
\newblock {Gorilla: A fast, scalable, in-memory time series database}.
\newblock {\em VLDB}, 2015.

\bibitem{seth2015granger}
A.~K. Seth, A.~B. Barrett, and L.~Barnett.
\newblock {Granger causality analysis in neuroscience and neuroimaging}.
\newblock {\em Journal of Neuroscience}, 2015.

\bibitem{shimizu2006linear}
S.~Shimizu, P.~O. Hoyer, A.~Hyv{\"a}rinen, and A.~Kerminen.
\newblock {A linear non-Gaussian acyclic model for causal discovery}.
\newblock {\em JMLR}, 2006.

\bibitem{spirtes2000causation}
P.~Spirtes, C.~N. Glymour, and R.~Scheines.
\newblock {Causation, prediction, and search}.
\newblock 2000.

\bibitem{tao2014metric}
C.~Tao, Y.~Ge, Q.~Song, Y.~Ge, and O.~A. Omitaomu.
\newblock {Metric ranking of invariant networks with belief propagation}.
\newblock {\em ICDM}, 2014.

\bibitem{tenenbaum2003theory}
J.~B. Tenenbaum and T.~L. Griffiths.
\newblock {Theory-based causal inference}.
\newblock {\em NIPS}, 2003.

\bibitem{tukey1977exploratory}
J.~W. Tukey.
\newblock {Exploratory data analysis}.
\newblock 1977.

\bibitem{wang2013theoretical}
Y.~Wang, L.~Wang, Y.~Li, D.~He, W.~Chen, and T.-Y. Liu.
\newblock {A theoretical analysis of NDCG ranking measures}.
\newblock {\em COLT}, 2013.

\bibitem{weisstein2004bonferroni}
E.~W. Weisstein.
\newblock {Bonferroni correction}.
\newblock 2004.

\bibitem{yang2014druid}
F.~Yang, E.~Tschetter, X.~L{\'e}aut{\'e}, N.~Ray, G.~Merlino, and D.~Ganguli.
\newblock {Druid: A real-time analytical data store}.
\newblock {\em SIGMOD}, 2014.

\bibitem{zaharia2016apache}
M.~Zaharia, R.~S. Xin, P.~Wendell, T.~Das, M.~Armbrust, A.~Dave, X.~Meng,
  J.~Rosen, S.~Venkataraman, M.~J. Franklin, et~al.
\newblock {Apache spark: a unified engine for big data processing}.
\newblock {\em CACM}, 2016.

\end{thebibliography}


\begin{thebibliography}{1}

\bibitem{bowman:reasoning}
M.~Bowman, S.~K. Debray, and L.~L. Peterson.
\newblock Reasoning about naming systems.
\newblock {\em ACM Trans. Program. Lang. Syst.}, 15(5):795--825, November 1993.

\bibitem{braams:babel}
J.~Braams.
\newblock Babel, a multilingual style-option system for use with latex's
  standard document styles.
\newblock {\em TUGboat}, 12(2):291--301, June 1991.

\bibitem{clark:pct}
M.~Clark.
\newblock Post congress tristesse.
\newblock In {\em TeX90 Conference Proceedings}, pages 84--89. TeX Users Group,
  March 1991.

\bibitem{herlihy:methodology}
M.~Herlihy.
\newblock A methodology for implementing highly concurrent data objects.
\newblock {\em ACM Trans. Program. Lang. Syst.}, 15(5):745--770, November 1993.

\bibitem{Lamport:LaTeX}
L.~Lamport.
\newblock {\em LaTeX User's Guide and Document Reference Manual}.
\newblock Addison-Wesley Publishing Company, Reading, Massachusetts, 1986.

\bibitem{salas:calculus}
S.~Salas and E.~Hille.
\newblock {\em Calculus: One and Several Variable}.
\newblock John Wiley and Sons, New York, 1978.

\end{thebibliography}
% You must have a proper ".bib" file
%  and remember to run:
% latex bibtex latex latex
% to resolve all references

\begin{appendix}
  %% \section{Intuition for predictability to measure causal relevance}
%% The framework of causal Bayesian Networks gives us an intuition for
%% reasoning about cause and effect, and quantify correlations between
%% pairs of variables in the presence of confounders.  Consider the DAG
%% X1 -> X2 -> X3 -> X4, which can be represented in a structural form as
%% follows,
%% \begin{align*}
%%   x_1 &= u_1,\\
%%   x_2 &= a_1 x_1 + u_2,\\
%%   x_3 &= a_2 x_2 + u_3,\\
%%   x_4 &= a_3 x_3 + u_4.
%% \end{align*}
%% where the dependencies between $X_i$s are captured by the linear
%% functions and the disturbances $u_i$ encode the observed/omitted
%% factors affecting $x_i$.  Consider the information that

\section{Dissecting the $r^2$ score: Controlling false positives}\label{sec:r2}
The goal in this section is to develop a systematic way of controlling
for false positives when testing multiple hypotheses.  Recall that a
false positive here means that \sys returns a hypothesis $(\X,\Y,\Z)$
in its top-$k$, implying that $\X \not\perp \Y \mid \Z$, when in fact
the alternate hypothesis that $\X\perp\Y\mid\Z$ is true.  We first
consider the ordinary least squares (OLS) scoring method to simplify
exposition.  Then, we show how \sys can adapt in a data-dependent way
to control false positives, and finally we conclude with future
directions to further improve the ranking.

\begin{figure*}[t]
\centering
\includegraphics[width=\textwidth]{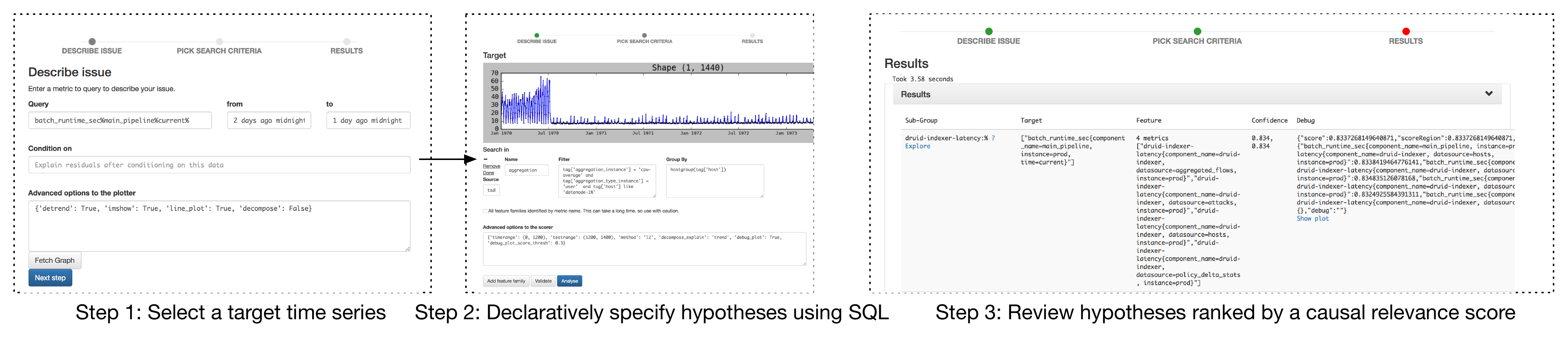}
\caption{Screenshots of \sys workflow for the end-user.}
\label{fig:explainit-webapp}
\end{figure*}

\subsection{The distribution of $r^2$ under the NULL}
Consider an OLS regression between features $\X$ of dimension $n
\times p$ ($n$ is the number of data points and $p$ is the number of
univariate predictors) and a target $\Y$ (for simplicity, of dimension
$n \times 1$), where we learn the parameters $\beta$ of dimension $p
\times 1$:\[
\Y = \X \beta + \epsilon
\]
where $\epsilon \sim \mathcal{N}(0, \sigma^2 \mathbb{I}_n)$ is an
error term; the distributional assumption on $\epsilon$ is convenient
for analysis.

The output of OLS is an estimate of $\beta$: $\hat{\beta}$ that
minimises the least squared error $\|\Y - \X \beta|_2^2$.  Let
$\hat{\Y} = \X \beta$ be the predicted values, and $(\Y-\hat{\Y})$ be
the residuals.  Define $r^2$ as follows:
\begin{align*}
  r^2 &\equiv 1 - \frac{\left(\Y - \hat{\Y}\right)^2}{\left(\Y-\bar{\Y}\right)^2} \\
      &= 1 - \frac{\rm RSS}{\rm TSS}
\end{align*}
where RSS is the Residual Sum of Squares, and TSS is the Total Sum of
Squares.  Notice that the TSS is computed after subtracting the mean
of the target variable $\Y$.  This means that the $r^2$ score compares
the predictive power of the linear model with $\X$ as its features, to
an alternate model that simply predicts the mean of the target
variable $\Y$.  The training and the mean are computed using the
training data.  Since the data $\Y$ is a {\em finite} sample drawn from the
distribution\[
\Y \mid \X \sim \mathcal{N}(\X \beta, \sigma^2 \mathbb{I}_n)
\]
any quantity (such as $\hat{\beta}$, $r^2$) computed from finite data
has a sampling distribution.  Knowing this sampling distribution can
be useful when interpreting the data, doing a statistical test, and
controlling false positives.

Under the hypothesis that there is no dependency between $\Y$ and
$\X$---i.e., the true coefficients $\beta = 0$---the sample statistic
$r^2$ is known~\cite{koerts1969theory,barten1962note,stackexchange-r2}
to be beta-distributed\[ r^2 \sim Beta\left (\frac {p-1}{2}, \frac
{n-p}{2}\right)
\]
The mean $\mu$ of this distribution is $(p-1)/(n-1)$, which tends to 1
as $p\rightarrow n$.  This conforms to the ``overfitting to the data''
intuition that when the number of predictors $p$ increase, $r^2
\rightarrow 1$ even when there is no dependency between $\Y$ and $\X$.
The distribution under the alternate hypothesis (that $\beta \neq 0$)
is more difficult to express in closed form and depends on the unknown
value $\beta$ for a given problem instance~\cite{cramer1987mean}.  The
variance of $r^2 \sim Beta(a,b)$ distribution is
\begin{align*}
  \operatorname{var}(r^2) &= \frac{a b}{(a+b)^2 (a+b+1)} \\
  &= \frac{\mu (1 - \mu)}{1+(n-1)/2} \\
  &\leq \frac{1}{4 (1 + (n-1)/2)} \\
  &= \mathcal{O}\left(\frac{1}{n}\right)
\end{align*}
So, we can see that the spread of the distribution around its mean
falls as $1/n$, as the number of data points $n$ increases.

\begin{figure}[t]
\centering
\includegraphics[width=0.45\textwidth]{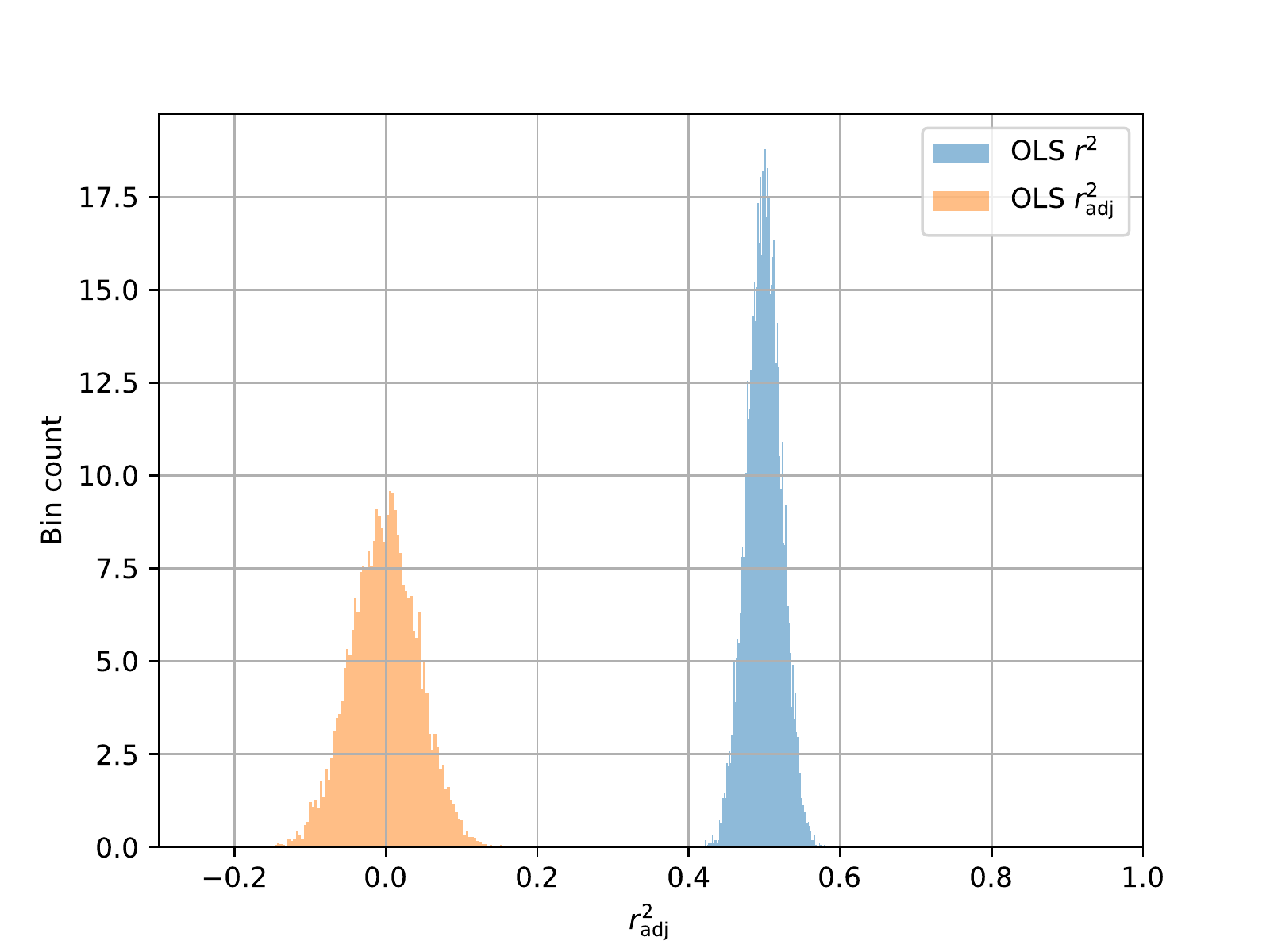}
\caption{A density plot of samples drawn from the distribution of
  $r^2$ and $r_{\rm adj}^2$ when $n=1000, p=500$, under the hypothesis
  that there is no relationship between $\X$ (of dimension $n \times
  p$) and a univariate $\Y$ (of dimension $n \times 1$).}
\label{fig:r2-adj-r2}
\end{figure}

To fix the over-fit problem, it is known that one can adjust $r^2$ for
the number of predictors using Wherry's
formula~\cite{olejnik2000using} to compute
\[ r_{\rm adj}^2 = 1 - (1 - r^2)\left(\frac{n-1}{n-p}\right)
\]
While it is difficult to compute the exact distribution of $r_{\rm
  adj}^2$, we can find that (under the hypothesis that there is no
dependency)
\begin{align*}
  \mathbb{E}[r_{\rm adj}^2] &= 0 \\
  \operatorname{var}[r_{\rm adj}^2] &= \left(\frac{2 (p-1)}{n-p}\right)\left(\frac{1}{n+1}\right)
\end{align*}
Notice that the variance increases as $p\rightarrow n$;
Figure~\ref{fig:r2-adj-r2} contrasts the two distributions empirically
for $n=1000, p=500$.

\ifarxiv
\begin{figure}[t]
\centering
\includegraphics[width=0.45\textwidth]{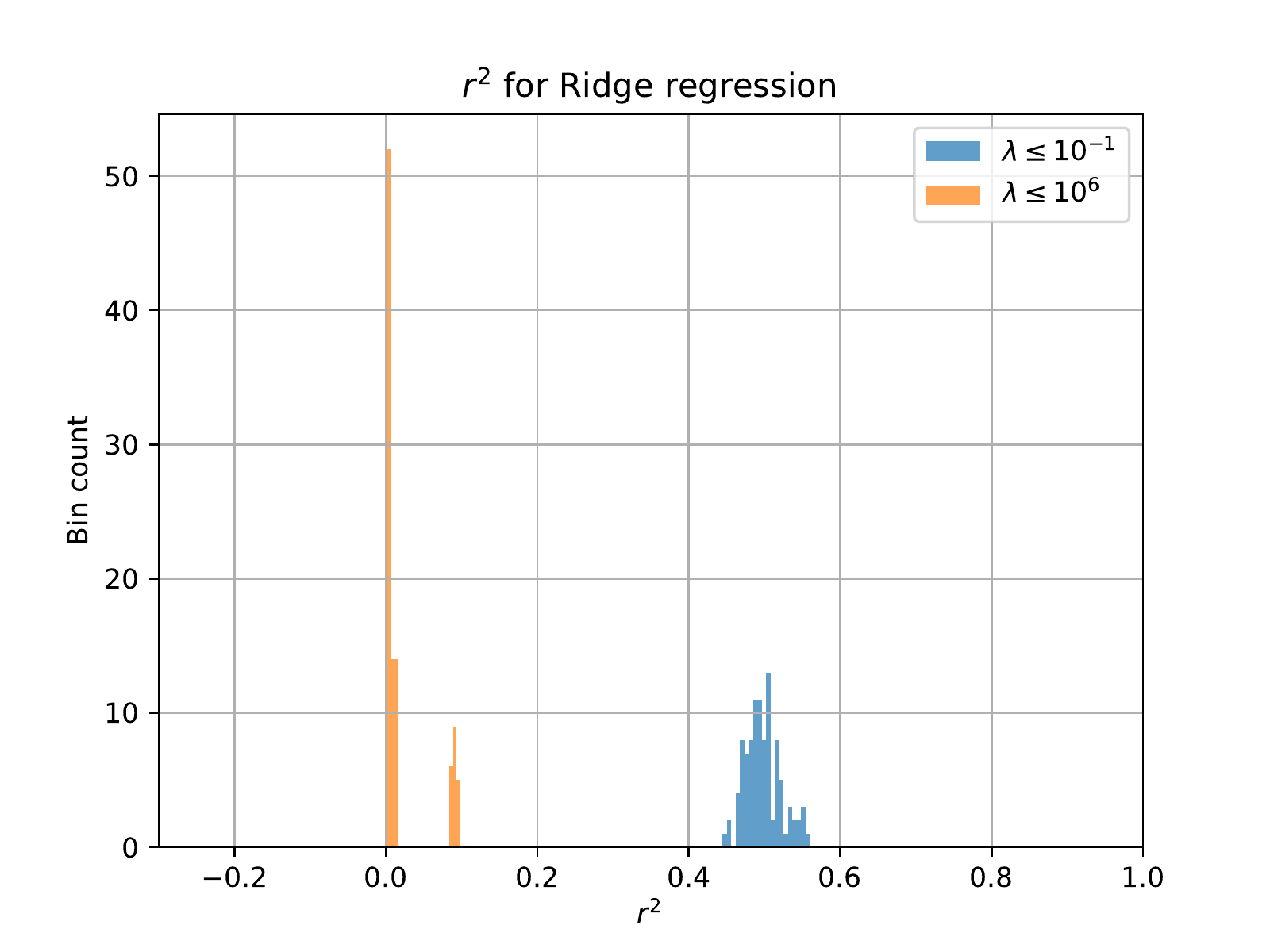}
\caption{The empirical density of $r^2$ calculated from sampling 100
  problem instances ($n=1000, p=500$) under the NULL hypothesis with
  univariate $\Y$.  We see that for a small $\lambda$, Ridge behaves
  similar to OLS's $r^2$. However, if we run Ridge regression with a
  grid search to select $\lambda$ using cross-validation, it selected
  $\lambda \approx 5 \times 10^{5}$ for which Ridge regression's
  empirical $r^2$ behaves more like $r_{\rm adj}^2$ of OLS (biased
  towards 0), but it also has a lower variance.  The bimodal behaviour
  arose because the regression chose two different values of $\lambda$
  in the samples across problem instances.}
\label{fig:r2-ridge}
\end{figure}
\fi

% What about ridge regression?
In \sys we use Ridge regression, which is harder to analyse than OLS.
However, we calculated the empirical distribution under the hypothesis
that there is no dependency between $\X$ and $\Y$, by sampling the
feature matrix $\X$ and $\Y$ whose entries are each drawn i.i.d@~from
$\mathcal{N}(0,1)$.  As we increased the ridge penalty parameter
$\lambda$ in the loss function ($T$ is the number of data points)\[
L_{\lambda}(\X,\Y) = \frac{1}{T}\|\Y - \X \beta\|_2^2 + \lambda
\|\beta\|_2^2
\]
and did model selection using cross-validation.  We find that $r^2$
value from Ridge regression behaved similar to the adjusted $r_{\rm
  adj}^2$ from OLS for the cross-validated $\lambda$, in the sense
that it tends towards the true value 0 under the NULL with a smaller
variance\ifarxiv (see Figure~\ref{fig:r2-ridge})\fi.  \ifarxiv\else In
fact, we can bound the variance of the score for Ridge and show that
it decreases monotonically with increasing regularisation strength
$\lambda$, and sample size $n$.  See our extended paper for more
details~\cite{explainit-extended}.\fi

\smallsec{Takeaways}: There are three takeaways from the above
analysis.  First, it highlights why the plain $r^2$ is biased towards
1 even when there is no relationship in the data and it is important
to adjust for the bias to get $r_{\rm adj}^2$.  Second, it shows that
$r_{\rm adj}^2$ is a sample statistic that has a mean and variance as
a function of the number of predictors $p$ and data points $n$.  In
the OLS case, we find that the variance drops as $\mathcal{O}(1/n)$,
where $n$ is the number of data points if the number of predictors $p
< n$ also increases linearly with $n$.  Third, although the analysis
does not directly applicable to Ridge regression, the cross-validated
$r^2$ statistic output by \sys behaves qualitatively like OLS's
$r_{\rm adj}^2$.

\subsection{False-positives: The $p$-value of the score}
The score output by \sys is equivalent to $r_{\rm adj}^2$ of OLS.
With knowledge about the mean and variance of the score, we can
approximate the $p$-value to each score $s$ to quantify: ``What is the
probability that a score at least as large as $s$ could have occurred
purely by chance, assuming the NULL hypothesis $H_0$ is true?''  This
quantity, $P(r_{\rm adj}^2 \geq s \mid H_0)$, can be estimated as
follows using Chebyshev's inequality (we drop $H_0$ for brevity):
\begin{align*}
  P(r_{\rm adj}^2 \geq s) &\leq \frac{\operatorname{var}{(r_{\rm adj}^2})}{s^2} \\
  &= \left(\frac{2(p-1)}{(n-p)(n-1)}\right)\frac{1}{s^2}
\end{align*}

Consider the scoring method $L_2-P50$, where there are 50 predictors.
If we have one day's worth of data at minute granularity ($n=1440$)
the $p$-value for a score $s$ can be approximated as $p(s) \approx 4.9
\times 10^{-5} / s^2$.  The inequality can be bounded more sharply
using higher moments of $r_{\rm adj}^2$ and higher powers of $s$, but
this approximation is sufficient to give us a few insights and help us
control false positives since \sys is scoring multiple hypotheses
simultaneously.

\smallsec{Controlling false-positives}: Given a ranking of scores
\linebreak $(s_1, \ldots, s_k)$ (in decreasing order) and their
corresponding $p$-values $(p_1(s_1), \ldots, p_k(s_k))$, we can
compute a new set of $p$-values using different techniques, such as
Boneferroni's correction~\cite{weisstein2004bonferroni} or
Benjamini-Hochberg~\cite{benjamini1995controlling} method, to declare
$l < k$ hypotheses ``statistically significant'' for further
investigation.  With the number of data points usually in the
thousands in our experiments, we find that the $p$-values for each
score are low enough that the top-20 results could not have occurred
purely by chance (assuming no dependency).  This is even after
applying the strict Boneferroni's correction for $p$-values, which
means that controlling for false-positives in the classical sense of
NULL-hypothesis significance testing is not much of a concern unless
the $r^2$ scores are very low; for instance when $s = 0.03$, the
$p$-value for $n=1000, p=50$ is $\approx 0.05$.

%%%%%%%%%%
%\begin{comment}
\ifarxiv
\smallsec{Ridge Regression} We outline an asymptotic argument for
Ridge regression for completeness, which is also used in cases where
$p \geq n$.  In general, it is difficult to compute the exact
distribution of the residual sum of squares (RSS) to obtain a bound on
its variance.  However, we can approximate it and show that its
variance has two properties: (1)~a similar asymptotic behaviour as
$r_{\rm adj}^2$ from OLS, and (2)~a monotonically decreasing function
of $\lambda$.  First, note that the solution to Ridge regression at a
specific regularisation strength $\lambda$ can be written as:
$\hat{\Y} = H \Y$, where $H = \X (\X^T\X + \lambda I)^{-1} \X^T$.
Then, RSS can be computed as follows:
\begin{align*}
  {\rm RSS} &= \| \Y - \hat{\Y} \|_2^2 \\
  &= \| (\mathbf{I} - H) \Y \|_2^2 \\
  &= \Y^T (\mathbf{I} - 2H + H^T H) \Y
\end{align*}

Under the NULL hypothesis, if $\Y \sim \mathcal{N}(0, \sigma^2 I)$,
the RSS is a quadratic of the form $\Y^T A \Y$, where $A = (I - 2H +
H^TH)$, and is distributed as ${\rm RSS} \sim
\chi_{\operatorname{trace}(A)}^2$.  It can be shown that the degrees
of freedom of this distribution can be written as:
\begin{align*}
  \operatorname{trace}(A)
  &= \operatorname{trace}(\mathbf{I} - 2H + H^T H) \\
  &= n - 2 \sum\limits_{j=1}^{p}\frac{d_j^2}{d_j^2 + \lambda} + \sum\limits_{j=1}^{p}\left(\frac{d_j^2}{d_j^2 + \lambda}\right)^2
\end{align*}
where $d_j^2$'s are the eigenvalues of $\X^T\X$.  Note that the trace
is a monotonically decreasing function of $\lambda$.

Similarly, we can work out that the total sum of squares TSS is
distributed as ${\rm TSS} \sim \chi_{n-1}^2$.  The score $r_{\rm
  adj}^2$ for Ridge regression is simply:
\begin{align*}
  r_{\rm adj}^2 &= 1 - \frac{\rm RSS}{\rm TSS} \\
  &= 1 - \frac{\epsilon^T \left(\mathbf{I} - 2H + H^T H\right) \epsilon}{\epsilon^T \left(\mathbf{I}-\frac{J}{n}\right)} \\
  &= \frac{\epsilon^T \left(2H - \frac{J}{n} - H^T H\right) \epsilon}{\epsilon^T \left(\mathbf{I}-\frac{J}{n}\right) \epsilon}
\end{align*}

To bound the variance of the fraction (call it $U/V$), we will use
proceed in three steps: First, for large $n$, we can invoke Central
Limit Theorem and show that $U$ and $V$ approach normal distributions:
$U \sim \mathcal{N}(\mu_u, 2\mu_u)$ and $V \sim \mathcal{N}(\mu_v,
2\mu_v)$.  Second, let us assume that the joint distribution of $U,V$
can be characterised by their means $\mu_u,\mu_v$, marginal variances
$\sigma_u,\sigma_v$ and some correlation coefficient $\rho$,
satisfying $-1 \leq \rho \leq 1$.  Third, we will use the fact that if
a random variable is bounded to an interval $[l,h]$, the variance is
$\leq (h-l)^2/4$.

To bound the variance of the fraction, we will consider typical values
of $V$, and identify a region where $U,V$ are most likely to be
jointly concentrated, and bound the variance in this region.  Since
$V$ is asymptotically normal, using Chernoff bounds, we can show:
\begin{align*}
  P(V \geq \mu_v - \gamma_v\sigma_v)
  &= 1 - P(V \leq \mu_v - \gamma_v\sqrt{2v}) \\
  &\geq 1 - e^{\BigO(\gamma_v^2)}
\end{align*}
Hence, $V$ marginally lies in this range with overwhelming
probability:
\begin{align*}
  V &\in [\mu_v-\gamma_v\sqrt{2\mu_v}, \mu_v+\gamma_v\sqrt{2\mu_v}]
\end{align*}
However, since $U$ and $V$ are not independent of each other, we
should consider the behaviour of $U\mid V=v$ in the ratio $U/V$.  For
any $V=v$, we can show that:
\begin{align*}
  U \mid (V=v) &\sim \mathcal{N}\left(\mu_u + \rho \frac{\sigma_u}{\sigma_v}(v-\mu_v), (1-\rho^2)\sigma_u^2\right)
\end{align*}
Hence, for any $V=v$ the mean shifts {\em linearly} in $v$ and the
variance is independent of $v$.  So, it is sufficient to consider the
behaviour of $U$ at the endpoints of the interval in which $V$ is most
likely to be concentrated.  When $V=V_{\rm min}=\mu_v-\gamma_v\sqrt{2\mu_v}$, we
have:
\begin{align*}
  U \mid \left(V=V_{\rm min}\right)
  &\sim \mathcal{N}\left(\mu_u+\rho\sqrt{\frac{\mu_u}{\mu_v}}(-\gamma_v\sqrt{2\mu_v}), 4(1-\rho^2)\mu_u)\right) \\
  &\sim \mathcal{N}\left(\mu_u-\rho\gamma_v\sqrt{2\mu_u}), 4(1-\rho^2)\mu_u\right)
\end{align*}
Notice that the $\sqrt{\mu_v}$ factor cancels out: that is, the range
of $U\mid V$ does not depend on the mean or variance of $V$ at all.
Also note that $U$ will be largest when $\rho < 0$, which agrees with
our intuition that $U/V$ will be large when $U$ and $V$ are negatively
correlated.  Thus, for typical values of $V$, $U\mid V$ will be
concentrated in the range:
\begin{align*}
  U \mid V &\in [\mu_u-(\rho+\BigO(\sqrt{1-\rho^2}))\BigO(\sqrt{\mu_u}), \\
           & \qquad \mu_u-(\rho-\BigO(\sqrt{1-\rho^2}))\BigO(\sqrt{\mu_u})]
\end{align*}
Conditional on $V$, we can see that $U$ lies in an interval of width
$\BigO(\sqrt{\mu_u})$.  Thus, the random variable $U/V$ lies in this
interval with high probability:
\begin{align*}
  \frac{U}{V} &\in \left[\frac{\mu_u-\BigO(\sqrt{\mu_u})}{\mu_v+\BigO(\sqrt{\mu_v})},
                    \frac{\mu_u+\BigO(\sqrt{\mu_u})}{\mu_v-\BigO(\sqrt{\mu_v})}\right]
\end{align*}
Therefore, the variance can be bounded by:
\begin{align*}
  \operatorname{var}\left[\frac{U}{V}\right]
  &\leq \frac{1}{4}\left(\frac{\mu_u+\BigO(\sqrt{\mu_u})}{\mu_v-\BigO(\sqrt{\mu_v})} -
  \frac{\mu_u-\BigO(\sqrt{\mu_u})}{\mu_v+\BigO(\sqrt{\mu_v})}\right)^2 \\
  &\approx O\left(\frac{\mu_u}{\mu_v^2}\right)
\end{align*}

Setting $U$ and $V$ appropriately, we can see that:
\begin{align*}
  \operatorname{var}(r_{\rm adj}^2)
  &= \operatorname{var}\left[\frac{\epsilon^T \left(2H - \operatorname{diag}(\frac{1}{n}) - H^T H\right) \epsilon}{\epsilon^T (\mathbf{I} - J/n) \epsilon}\right] \\
  &= \BigO\left(\frac{{\rm df}}{(n-1)^2}\right)
\end{align*}
where the effective degrees of freedom ${\rm df}$ is the trace of the
numerator:
\begin{align*}
  {\rm df} &= \sum\limits_{j=1}^{p}\left(\frac{2 d_j^2}{d_j^2 + \lambda} - \frac{1}{n} - \left(\frac{d_j^2}{d_j^2 + \lambda}\right)^2\right)
\end{align*}

Here, $d_j^2$ are the eigenvalues of $X^T X$, and $p$ is the number of
features.  Note that the effective degrees of freedom is also
monotonically decreasing with higher $\lambda$, and can be
approximated in a data-dependent fashion.  As $\lambda\rightarrow 0$,
${\rm df}\rightarrow p-1$ (OLS case), and as
$\lambda\rightarrow\infty$, ${\rm df}\rightarrow 0$, and $r_{\rm
  adj}^2\rightarrow 0$.  Moreover, it does not depend on the variance
$\sigma^2$.
%\end{comment}
\fi

%% \section{Why joint models?}\label{sec:why-joint}
%% Joint models have the advantage that they can model more complex
%% relationships in data, but as we saw in the previous section they are
%% also prone to overfitting, and have higher computational cost.  In
%% this section, we discuss standard examples that illustrate certain
%% non-intuitive behaviour (at first glance).

%% Can a univariate variable be uncorrelated with the target, but still
%% be useful when taken jointly with other variables?

%% \section{Conditioning and ranking: How bad can it be?}\label{sec:cond-ranking}
%% Compared to existing tools, \sys is unique in the sense that it
%% exposes conditioning as an operator to control for simultaneous
%% sources of variation that can occur in the system, to improve ranking.
%% As we saw in~\label{subsec:conditioning}, conditioning can drastically
%% improve ranking and surface diverse causes to the user.  In this
%% section, we give an intuitive argument that shows

%\subsection{Controlling false positives}\label{subsec:fdr}

\ifarxiv
\section{Correctness of the conditional regression procedure}\label{sec:cond-proof}
In~\S\ref{subsec:impl:scoring}, we used a procedure to score to what
extent $\X \perp \Y \mid \Z$, where a zero score means $\X \perp \Y
\mid \Z$.  We provide a proof of this standard procedure: if
$\X,\Y,\Z$ are jointly multivariate normally distributed, then a zero
score is equivalent to stating that $\X \perp \Y \mid \Z$.

Without loss of generality, we assume that the variables $\X,
\Y, \Z$ are centred so their mean is $\mathbf{0}$.  If $(\X,\Y,\Z)
\sim \mathcal{N}(\mathbf{0}, \boldsymbol\Sigma)$, where the covariance
matrix $\boldsymbol\Sigma$ is partitioned into the following block
matrices
\begin{align*}
  \boldsymbol\Sigma
  &= \mathbf{E}[\begin{bmatrix}\X \; \Y \; \Z\end{bmatrix}
      \begin{bmatrix}\X \; \Y \; \Z\end{bmatrix}^T] \\
  &=
\begin{bmatrix}
 \boldsymbol\Sigma_{xx} & \boldsymbol\Sigma_{xy} & \boldsymbol\Sigma_{xz} \\
 \boldsymbol\Sigma_{yx} & \boldsymbol\Sigma_{yy} & \boldsymbol\Sigma_{yz} \\
 \boldsymbol\Sigma_{zx} & \boldsymbol\Sigma_{zy} & \boldsymbol\Sigma_{zz} \\
\end{bmatrix}
\end{align*}
then the conditional variance $\boldsymbol\Sigma_{xy;z}$ can be
written as~\cite{eaton1983multivariate,wikiconditional}:
\begin{align*}
  \boldsymbol\Sigma_{xy;z}
  &= \begin{bmatrix}
    \boldsymbol\Sigma_{xx} & \boldsymbol\Sigma_{xy} \\
    \boldsymbol\Sigma_{yx} & \boldsymbol\Sigma_{yy}
  \end{bmatrix} - \begin{bmatrix}
    \boldsymbol\Sigma_{xz} \\
    \boldsymbol\Sigma_{yz}
  \end{bmatrix}
  \boldsymbol\Sigma_{zz}^{-1}
  \begin{bmatrix}
    \boldsymbol\Sigma_{xz} & \boldsymbol\Sigma_{yz}
  \end{bmatrix} \\
 &= \begin{bmatrix}
    \ldots & \boldsymbol\Sigma_{xy}-\boldsymbol\Sigma_{xz}\boldsymbol\Sigma_{zz}^{-1}\boldsymbol\Sigma_{zy} \\
    (\boldsymbol\Sigma_{xy}-\boldsymbol\Sigma_{xz}\boldsymbol\Sigma_{zz}^{-1}\boldsymbol\Sigma_{zy})^T & \ldots
  \end{bmatrix}
\end{align*}

Hence, to prove that $\X \perp \Y \mid \Z$, we just need to show that
the off-diagonal entry---the cross-covariance matrix between $\X$ and
$\Y$ conditional on $\Z$---i.e.,
$\boldsymbol\Sigma_{xy}-\boldsymbol\Sigma_{xz}\boldsymbol\Sigma_{zz}^{-1}\boldsymbol\Sigma_{zy}
= \mathbf{0}$.  Now recall that the procedure involves three
regressions:
\begin{enumerate}
\item $\X \sim \Z$, with predictions $\hat{\X}$ and residuals
  $R_{\X;\Z}$,
\item $\Y \sim \Z$, with predictions $\hat{\Y}$ and residuals
  $R_{\Y;\Z}$,
\item $R_{\X;\Z} \sim R_{\Y;\Z}$, with residuals $R_{\X,\Y,\Z}$, and
  the score being the $r^2$ of this final regression.
\end{enumerate}

Consider the regression $\X \sim \Z$, which denotes $\X = \beta_x \Z +
\epsilon$, whose solution $\beta_x$ is the minimiser of the squared
loss function $\|\X - \beta_x \Z\|_F^2$.\footnote{Since $\X$ is a
  matrix, $\|\X\|_F^2 = \sum X_{ij}^2$.} It can be shown by
differentiating the loss with respect to $\beta_x$ that the solution
is the matrix:
\begin{align*}
  \beta_x &= \X \Z^T (\Z \Z^T)^{-1}
\end{align*}

Hence, the residuals $R_{\X;\Z}$ (and similarly, $R_{\Y;\Z}$) can be
written as:
\begin{align*}
  R_{\X;\Z} &= \X - \beta_x \Z \\
  &= \X - \X \Z^T (\Z \Z^T)^{-1} \Z \\
  R_{\Y;\Z} &= \Y - \beta_y \Z \\
  &= \Y - \Y \Z^T (\Z \Z^T)^{-1} \Z
\end{align*}

Now, consider the geometry of the third OLS regression, $R_{\X;\Z}
\sim R_{\Y;\Z}$, whose score is the one \sys returns.  A zero (low)
score means there is no (low) explanatory power in this regression.
Since the OLS regression considers linear combinations of the
independent variable ($R_{\Y;\Z}$), consider what happens if we view
the dependent and independent variables as vectors: a zero score can
happen only when the dependent and independent variables are {\em
  orthogonal} to each other.  That is,
\begin{align}
  R_{\X;\Z} R_{\Y;\Z}^T &= \mathbf{0}\label{eqn:0}
\end{align}

Substituting the values in the above equation and expanding, we get:
\begin{align*}
  R_{\X;\Z} R_{\Y;\Z}^T &= (\X - \beta_x \Z) (\Y - \beta_y \Z)^T \\
  &= \X \Y^T - \X \Z^T \beta_y^T - \beta_x \Z \Y^T + \beta_x \Z \Z^T \beta_y^T
\end{align*}

Consider the last term in the product, and substitute the values for
$\beta_x$ and $\beta_y$ in it using the identity that $(A^{-1})^T =
(A^T)^{-1}$, and $(A B)^T = B^T A^T$, we have:
\begin{align*}
  \beta_x \Z \Z^T \beta_y^T
  &= \beta_x (\Z \Z^T) \left(\Y \Z^T (\Z \Z^T)^{-1}\right)^T \\
  &= \beta_x (\Z \Z^T) \left((\Z \Z^T)^{-1}\right)^T \Z \Y^T \\
  &= \beta_x (\Z \Z^T) (\Z \Z^T)^{-1} \Z \Y^T \\
  &= \beta_x \Z \Y^T
\end{align*}

Hence, we can see that the dot product between the residuals
simplifies to:
\begin{align*}
  R_{\X;\Z} R_{\Y;\Z}^T
  &= \X \Y^T - \X \Z^T \beta_y^T - \beta_x \Z \Y^T + \beta_x \Z \Z^T \beta_y^T \\
  &= \X \Y^T - \X \Z^T \beta_y^T \\
  &= \X \Y^T - \X \Z^T (\Z \Z^T)^{-1} \Z \Y^T
\end{align*}

From equation~\ref{eqn:0}, we know that: $\X \Y^T - \X \Z^T (\Z
\Z^T)^{-1} \Z \Y^T = \mathbf{0}$.  The first term $\X \Y^T$ is a
sample estimate of the population covariance $\boldsymbol\Sigma_{xy}$.
Using that fact, we can get the desired result:
\begin{align*}
  \underbrace{\left(\X \Y^T\right)}_{\boldsymbol\Sigma_{xy}}
  -
  \underbrace{\left(\X \Z^T\right)}_{\boldsymbol\Sigma_{xz}}
  \underbrace{\left((\Z \Z^T)^{-1}\right)}_{\boldsymbol\Sigma_{zz}^{-1}}
  \underbrace{\left(\Z \Y^T\right)}_{\boldsymbol\Sigma_{zy}} &= \mathbf{0}
\end{align*}\qed
\fi

%% For the longer version of this paper.
\section{Example SQL queries}\label{sec:sql-queries}
In \sys, the user writes SQL queries at three phases: (1)~prepare data
for the target metric family ($\Y$), (2)~constrain the search space of
hypotheses from various data sources ($\X$), and (3)~a set of
variables to condition on ($\Z$).  The results from each phase is then
used to construct the hypothesis table using a simple join (in
Figure~\ref{fig:pipeline}).  We provide examples for SQL queries in
each phase that we used to diagnose issues in the case studies listed
in~\S\ref{sec:case-study}.  The tables used in these queries have more
features than listed below.

First, the user writes a query to identify the target metric that they
wish to explain.  In our implementation, we wrote an external data
connector to interface to expose data in our OpenTSDB-based monitoring
system to Spark SQL in the table {\tt tsdb}.  The schema for the table
is simple: each row has a timestamp column (epoch minute), a metric
name, a map of key-value tags, and a value denoting the snapshot of
the metric.  This result is stored in a temporary table tied to the
interactive workflow session with the user; here, we will refer to it
as {\tt Target} in subsequent queries.

\begin{lstlisting}[language=SQL,caption={Taget metric family},captionpos=b,label=Query]
SELECT
  timestamp, tag['pipeline_name'],
  AVG(value) as runtime_sec
FROM tsdb
WHERE metric_name = 'pipeline_runtime'
AND timestamp BETWEEN T1 and T2
GROUP BY
timestamp, tag['pipeline_name']
ORDER BY timestamp ASC
\end{lstlisting}

Next, the user specifies multiple queries to narrow down the feature
families.  We list network, and process-level features below.  The
{\tt flow} and {\tt processes} tables in these queries are from
another time series monitoring system.

% Queries to prepare data into feature families.  Examples from
% various data sources/connectors.

\begin{lstlisting}[language=SQL,caption={Network features},captionpos=b,label=Query]
SELECT
  timestamp, CONCAT(src_address, service_port),
  AVG(pkts), AVG(bytes),
  AVG(network_latency), AVG(retransmissions),
  AVG(handshake_latency), AVG(burstiness)
FROM flows
WHERE timestamp BETWEEN T1 and T2
GROUP BY timestamp, CONCAT(src_address, dst_port)
ORDER BY timestamp ASC
\end{lstlisting}

The above query produces 6 network performance features for every
source IP address, for every service that it talks to (identified by
service port), for every timestamp (granularity is minutes).  The
second stage in Figure~\ref{fig:pipeline} interprets the 6 aggregated
columns (pkts, bytes, network latency, retransmissions, and
burstiness) as a map whose keys are the column names, and values are
the aggregates.  Hence, we can union results from multiple queries
even though they have different number of columns in their schema.

\begin{lstlisting}[language=SQL,caption={Process-level features},captionpos=b,label=Query]
SELECT
  timestamp,
  CONCAT(service_name, SPLIT(hostname, '-')[0]),
  AVG(stime+utime) as cpu,
  AVG(statm_resident) as mem,
  AVG(read_b)
  AVG(greatest(write_b-cancelled_write_b,0)),
FROM processes
WHERE
  SPLIT(hostname,'-')[0] IN
  ('web', 'app', 'db', 'pipeline') AND
  timestamp BETWEEN T1 and T2
GROUP BY
timestamp,
CONCAT(service_name, SPLIT(hostname, '-')[0])
ORDER BY timestamp ASC
\end{lstlisting}

The above query also illustrates how we can group hostnames that are
numbered (e.g., {\tt web-1, web-2, ..., app-1, ...} etc.) into
meaningful groups ({\tt web, app}).  Enterprises typically have an
inventory database containing useful machine attributes such as the
datacentre, network fabric, and even rack-level information with every
hostname.  This side information can be used by joining on a key such
as the hostname or IP address of the machine.

The use of SQL also opens up more possibilities:
\begin{itemize}
  \item User-defined functions (UDFs) for common operations.  For
    example, we define a {\tt hostgroup} UDF instead of {\tt
      SPLIT(hostname, '-')[0]}.
  \item Windowing functions allow users to look back or look ahead in
    the time series to do smoothening and running averages.
  \item Ranking functions, such as percentiles, allow us to compute
    histograms that can be used to identify outliers.  For example, in
    a distributed service, the 99th percentile latency across a set of
    servers is often a useful performance indicator.
  \item Commonly used feature family aggregates (such as $99^{\rm th}$
    percentile latency) can be made available as materialised views to
    avoid expensive aggregations.
\end{itemize}

Finally, the user specifies a query to generate a list of variables to
condition on.  Here, we would like to condition on the total number of
input events to the respective pipelines.  This result is stored in a
temporary table called {\tt Condition}.

\begin{lstlisting}[language=SQL,caption={Conditioning variables},captionpos=b,label=Query]
SELECT
  timestamp, tag['pipeline_name'],
  AVG(value) as input_events
FROM tsdb
WHERE
  metric_name = 'pipeline_input_rate' AND
  timestamp BETWEEN T1 and T2
GROUP BY
timestamp, tag['pipeline_name']
ORDER BY timestamp ASC
\end{lstlisting}

% Queries to specify the hypothesis space.
\smallsec{Generating hypotheses}: Next, \sys generates hypotheses by
automatically writing join queries in the backend.  With SQL, \sys
also has the flexibility to impose conditions on the join to ensure
additional constraints on the join operation, which we show in the
example below.  Let $FF_i$ denote the resulting tables from the
feature family queries listed above, after transforming them into the
following normalised schema:

\begin{verbatim}
  timestamp: datetime
  name: string
  value: map<string, double>
\end{verbatim}

Next, \sys runs the following query to generate all hypotheses.  Note
that the inputs to the pipelines are matched correctly in the JOIN
condition.  We use {\tt X...} for brevity to avoid listing all
columns, but highlight the ordering of variables: Features ($\X$,
Target ($\Y$), Conditioning ($\Z$).

\begin{lstlisting}[language=SQL,caption={Generating hypotheses},captionpos=b,label=Query]
SELECT
  timestamp, X..., Y..., Z...
FROM
  (FF_1 UNION FF_2 UNION ... FF_n) FF
FULL OUTER JOIN
  Target ON
  (FF.timestamp = Target.timestamp)
FULL OUTER JOIN
  Condition ON
  Target.timestamp = Condition.timestamp AND
  Target.pipeline_name = Condition.pipeline_name
ORDER BY timestamp ASC
\end{lstlisting}

The result from this query is a multidimensional time series that is
then used by \sys for ranking.  The join type dictates the policy for
missing observations for the time series.  At this stage, \sys
optimises the representation into dense numpy arrays, scores each
hypothesis, and returns the top 20 results to the user.  Missing
values in the time series are interpolated to the closest non-null
observation.

% Queries that do the ranking.

  \ifarxiv
  \section{Lessons Learnt}\label{sec:lessons-learnt}
In this section we chronicle some important observations that we
learnt from our experience.

% results only as good as your data, and as also only as good as the
% hypotheses you ask.  Quite challenging with limited data, especially
% latent issues that show up over time.  E.g., slow memory leaks.
\begin{comment}
\smallsec{Adequacy of data}: Since \sys relies primarily on observed
data, it follows that the results are only as good as the quality of
data.  It is important for the data input to \sys to have instances of
both good and bad behaviour: trying to explain a ``high'' pipeline
runtime without enough data for what constitutes an acceptable runtime
did not work.  There were a handful of incidents where we could not
use \sys as we did not have data for good behaviour in the data.
\end{comment}

% Visualisation works: no true zeros or independencies.  Always have
% correlated variables to varying degrees.  thus, visualising the
% goodness of fit is handy in seeing how the predictions work.
\smallsec{Visualisations are important}: We found substantial benefits
in adding diagnostic plots to the results output by \sys, primarily to
diagnose errors in \sys, and also as a visual aid to the operator for
instances where a single confidence score is not adequate.  When
scoring $\X$ against $\Y$ conditioned on $\Z$, we show two plots for
every $\X$: the time series $\Y \mid \Z$ and the predicted value
$\operatorname{E}[\Y \mid \X, \Z]$ (e.g.,
Figure~\ref{fig:perf-residuals}).  This helped us draw conclusions and
instill confidence in our approach of using data to reason about
system performance.  For example, Figure~\ref{fig:good-vs-bad-spike}
shows how \sys is unable to explain the spike in the blue time series,
but its confidence in explaining the saw-tooth behaviour is high.

The case study in~\S\ref{subsec:conditioning} was another instance
where visualisations helped build confidence in the ranking: let $Y_r$
denote the runtime after conditioning on input size.  In
Figure~\ref{fig:perf-residuals} the blue plot shows $Y_r$, and the
green plot shows $\operatorname{E}[Y_r \mid \mathbf{X}]$, where
$\mathbf{X}$ is the feature family denoting packet retransmissions.
We see that the spikes in $Y_r$ that are above the mean are explained
by $\mathbf{X}$, but the spikes below the mean are not.  This is
interesting because it says that retransmissions explain increases in
runtimes, but not dips.

\smallsec{Attributing metadata}: In our experience, we found that
systems troubleshooting is useful only if the outcome is constructive
and actionable.  Thus, it is important to identify the key owners of
metrics and services, and it is important for them to understand what
the metrics mean.  For example, we find that broad infrastructure
metrics such as ``percent CPU utilisation'' are not useful unless the
CPU utilisation can be attributed to a service that can then be
investigated.  Fortunately, this arises naturally, as many of the
metrics we see in our data are published by individual services.

% Future: Reusing data and cautioning against garden of forking paths.
% Thresholdout is promising.
%% We are continuing to enhance \sys in the following ways: First, we are
%% formalising \sys's query language by combining it with a familiar
%% relational query language: SQL.  This combination is natural because
%% in \sys, a user expresses her search criteria for families in terms of
%% a ``filter,'' and ``groups'' families based on a shared attribute
%% prior to scoring.  The declarative nature of SQL is well suited for
%% expressing adhoc queries, feature transformations (e.g., counting the
%% number of occurrences of the string {\tt Exception} in log messages),
%% leaving the task of optimising (e.g., picking an appropriate scoring
%% function) to a runtime system.  Second, we are also working on
%% incorporating log messages, and other time series of semi-structured
%% data into \sys.  And finally, we are investigating approaches such as
%% Thresholdout~\cite{dwork2015reusable} to guard the end user from
%% overusing limited data to draw conclusions.
%%%%%%%%%%%%%%%%%%%%%
% Next set of papers: A database system to formalise such ``queries''
% to the system.  SQL, adhoc, etc.
%%%
% formalising the group approach in this paper, stats test, etc.

\begin{figure}[t]
\centering
\includegraphics[width=0.3\textwidth]{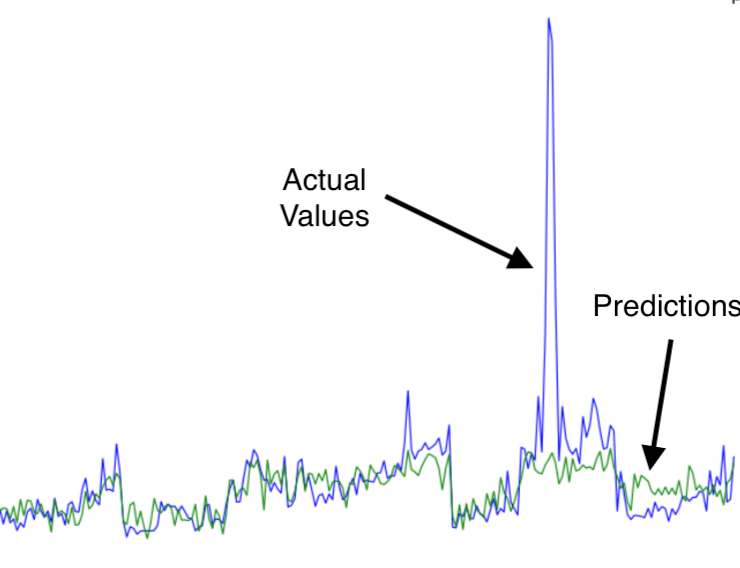}
\caption{The blue plot is our target runtime $Y$, and the green plot
  is the predicted values $E[Y \mid X]$ using CPU temperature values.
  Short of a precise loss function, a single score does not
  distinguish a good from a bad prediction.  Visualisations come in
  handy to rule out such explanations.}\vspace{-1em}
\label{fig:good-vs-bad-spike}
\end{figure}

% time series of text, and semi-structured data.
\begin{figure}[t]
\centering
\includegraphics[width=0.5\textwidth]{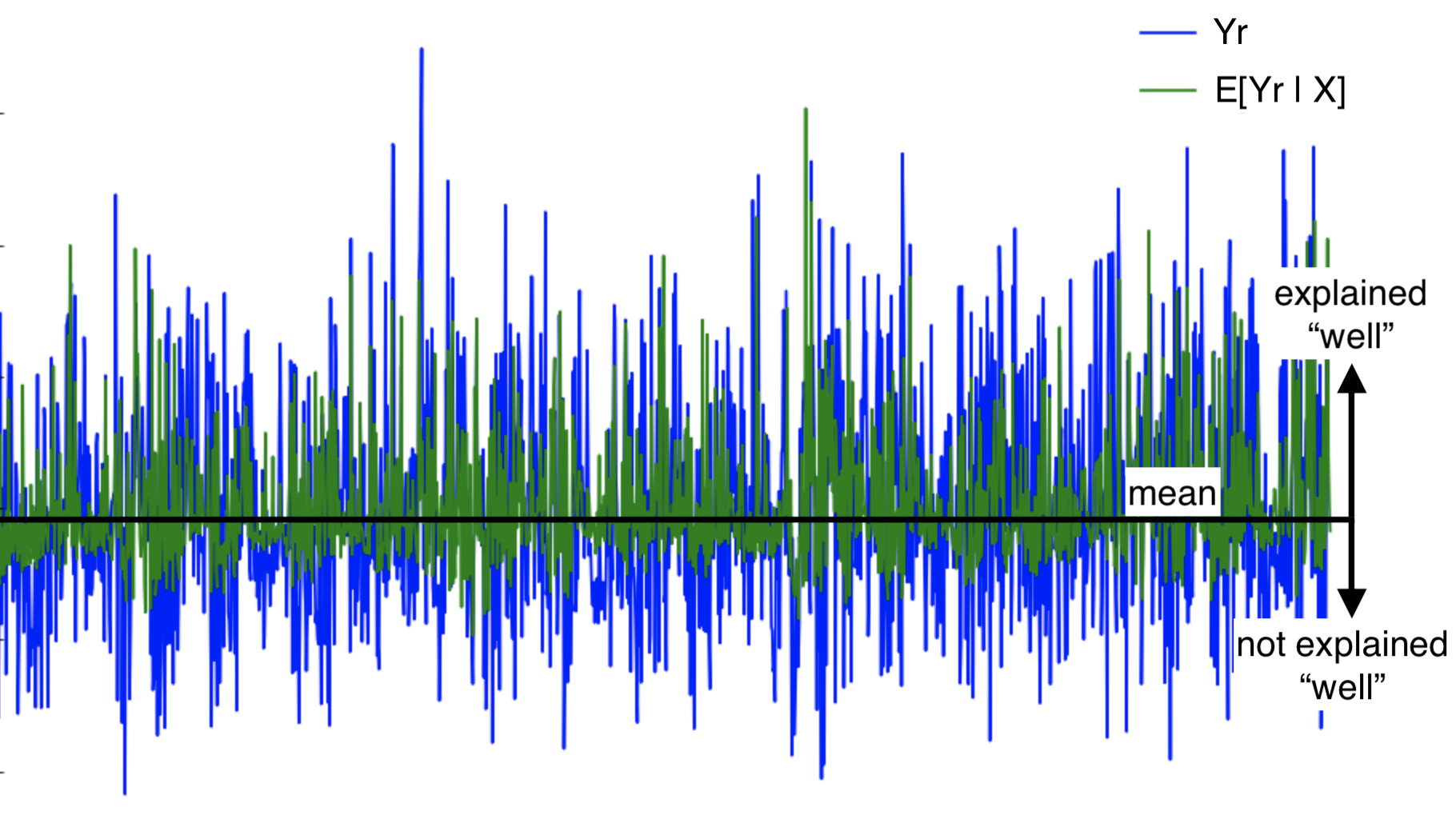}
\caption{The time series plot shows how spikes above the mean are well
  explained by packet retransmissions, whereas variations below the
  mean are not explained. (Best viewed in colour.)}
\label{fig:perf-residuals}
\end{figure}

  \fi
\end{appendix}

\end{document}